\pdfoutput=1

\documentclass[manuscript]{aastex62}

\graphicspath{{./}{figures/}}

\received{March 9, 2020}
\revised{August 24, 2020}
\accepted{September 7, 2020}
\submitjournal{ApJ}

\usepackage{booktabs}
\usepackage{color}
\usepackage{amssymb}
\usepackage[]{graphicx}
\usepackage{capt-of}
\usepackage{mathtools}
\usepackage{textcomp}
\usepackage{blindtext}
\usepackage{multirow}
\usepackage{floatrow}

\begin{document}
\title{\textbf{Subsecond Spikes in Fermi GBM X-ray Flux as a Probe for Solar Flare Particle Acceleration}}

\author{ Trevor Knuth }
\affil{ The School of Physics and Astronomy, University of Minnesota }

\author{ Lindsay Glesener }
\affil{ The School of Physics and Astronomy, University of Minnesota }

\begin{abstract}

Solar flares are known to release a large amount of energy into accelerating electrons. Studying small timescale ($\leq$ 2s) fluctuations in nonthermal X-ray flux offers the opportunity to probe the nature of those acceleration mechanisms. By comparing the durations, differences in timing between energy bands, and the periodicity of these spikes against the relevant timescales called for by various acceleration mechanisms, a test for each mechanism's validity can be made. This work details the analysis of fast fluctuations in \textit{Fermi} Gamma-ray Burst Monitor (\textit{Fermi} GBM) data from 2 M9.3 class solar flares that occurred on SOL2011-07-30 and SOL2011-08-04. This study shows the usefulness of \textit{Fermi} GBM data as a means of examining these small timescale spikes and presents a rigorous method of identifying, counting, and measuring the temporal properties of these subsecond X-ray spikes. In the 2 flares examined we found spikes to primarily occur in spans of 60-100 seconds in the impulsive phase. The relative spike intensity averaged between 6\% to 28\% when compared to the slowly varying component of the X-ray flux. The average spike durations were 0.49 and 0.38 seconds for the 2 flares. The spike duration distribution for the SOL2011-08-04 flare was found to follow a power law with a -1.2 $\pm$ 0.3 index. Of the 3 spiking intervals identified, only 1 was found to have a periodicity, showing significant power at the 1.7 $\pm$ 0.1 Hz frequency.

\end{abstract}

\section{Introduction}\label{sec:intro}
Solar flares release magnetic energy into at least 4 forms: radiation, plasma heating, magnetohydrodynamic (MHD) waves, and accelerated particle distributions. The details of particle acceleration in solar flares have been the topic of much debate over the years as an estimated 10\%-50\% \citep[e.g.][]{emslie} of the flare's energy is released as accelerated particles. The question of which mechanisms are at play is largely unsettled \citep{zharkova}. Popular theories include various invocations of first-order and/or second-order Fermi acceleration \citep[e.g.][]{tsuneta,chen}. First-order Fermi acceleration is the energization of particles that are, for example, rebounding off of turbulent fields near a shock front and generally requires an already energized population of particles to efficiently accelerate \citep{bell}. First-order Fermi acceleration can also be achieved in indirect means such as in the collapse of magnetic islands \citep[e.g.][]{drake, guidoni}. Second-order Fermi acceleration is the stochastic acceleration of particles off of randomly moving magnetic mirrors (such as in a magnetized cloud of plasma), with oncoming collisions accelerating particles and receding collisions decelerating particles \citep{fermi}. Net energization is achieved through the difference in the frequency between these 2 types of collisions. There also exist more exotic acceleration models such as ones that employ energy transport via Alfv\'{e}n waves \citep{fletcher} and simpler methods that invoke direct Lorentz acceleration \citep{zharkova}.

These energized particles can be studied via their high-energy emission. Electron collisions with ambient ions create nonthermal bremsstrahlung radiation. Through the use of thick- and thin-target emission models, the energetics of the particles can be reproduced from this emission \citep{brown}. It then follows that any structure in the time profiles of this emission will directly relate to the particles themselves, convolved with any significant propagation effects. Previous studies have positively identified sub-second spikes in hard X-ray ($>$ 10 keV) flux \citep{kiplinger1983,kiplinger1984,aschwanden1995,qiu,cheng,glesener,alty}. Comparing the temporal characteristics of these spikes against the temporal characteristics of acceleration mechanisms is a powerful tool in constraining the validity of various mechanisms.

\cite{kiplinger1983, kiplinger1984} studied fast spikes in hard X-ray solar flares observed by the \textit{Solar Maximum Mission's} hard X-ray burst spectrometer (\textit{SMM}-HXRBS) \citep{orwig}. This study found spikes with a full-width half-maximum as low as 45 ms, greatly constraining acceleration timescales of nonthermal acceleration models. They found that 10\% of the flares examined exhibited these subsecond spikes. However, the HXRBS only provided a 15-channel energy spectrum, offering limited spectral information. 

\cite{aschwanden1995,aschwanden} studied solar flare hard X-ray data from the Burst And Transient Source Experiment (BATSE) onboard the \textit{Compton Gamma-Ray Observatory} (\textit{CGRO}). These studies focused on localizing the region of acceleration via time-of-flight analysis and proposed a model based upon a dynamic loss cone angle to explain these subsecond spikes. An exponential distribution of spikes in the 0.3-1s range was discovered with the conclusion that with sufficiently large collection area, subsecond spikes could be observed in every flare. As BATSE was designed for faint astrophysical sources, it had a large collection area of 2025 cm$^{2}$ per detector. This would also lead to large pulse pileup issues for solar flares, limiting spectral sensitivity. 

\cite{qiu} and \cite{cheng} performed studies of fast time variations in hard X-ray data from the \textit{Reuvan Ramaty High Energy Solar Spectrometer} (\textit{RHESSI}). These studies were focused on locating the emission source of these spikes using \textit{RHESSI}'s imaging capability. The studies found that spikes were predominantly emitted from the footpoints and their presence was highly dependent on the active region creating the flares. The vast majority of their spikes originated from flares in only a few active regions. They also found their spike duration distribution to be closer to a power law, rather than an exponential distribution. As \textit{RHESSI} used a Fourier imaging method \citep{hurford} which utilizes a highly modulated time profile, a demodulation of the data was required \citep{arzner}. This data demodulation by necessity adds noise and thus decreases the sensitivity for the detection of fast variations.

\cite{alty} studied the fast time emission in the SOL2011-08-04 flare in multiple wavelengths, using data from the Nobeyama Radioheliograph \citep{nobeyama} as well as the \textit{RHESSI}, \textit{Fermi}, and \textit{WIND} spacecrafts \citep{aptekar}. This study found subsecond variations on the scale of 50 ms and concluded super-Dreicer electric field to be the primary method of acceleration.

These studies all performed remarkable science, but the instruments used all possessed distinct limitations. A modern alternative is the Gamma-ray Burst Monitor (GBM) onboard the \textit{Fermi Gamma-ray Space Telescope} \citep{meegan}. The GBM consists of an array of 12 thallium doped sodium-iodide detectors (NaI(Tl)) numbered n1 through n12 and placed in ``all-sky" arrangement. These detectors effectively cover an energy range of 8 keV to 1 MeV, with each detector possessing approximately 127 $\textrm{cm}^2$ of surface area. The primary issue with using \textit{Fermi} GBM is its lack of solar optimization. This can lead to substantial pulse pileup during large solar flares, complicating spectral analysis. As the surface area is substantially smaller than that of BATSE and the system can handle higher overall count rates, the problem of pileup is lessened (though not negligible). As the instrument does not utilize rotational modulation collimation, there is no modulated time signal. Finally, as the instrument was launched in 2008 it has a large observational overlap with \textit{RHESSI} and the \textit{Solar Dynamics Obsevatory} (\textit{SDO}), creating multi-instrument observations of many single events. All of these factors make the \textit{Fermi} GBM a capable instrument for timing studies. 

In this study we examine subsecond hard X-ray spikes in two solar flares using \textit{Fermi} GBM data and present a variety of analysis methods which can serve as a diagnostic tool for examining particle acceleration in solar flares. This study also examines the SOL2011-08-04 flare studied by \cite{alty} but focuses on the \textit{Fermi} GBM data and utilizing a variety of time domain analyses to extract more information about the subsecond spikes. It also applies these methods to another flare on SOL2011-07-30. Section \ref{sec:selection} will detail the flare selection. Section \ref{sec:identify} presents the method used to identify the spikes in these flares. Our method of fitting and determining the spike parameters is shown in Section \ref{sec:gmm}. The results of these analyses are presented in Section \ref{sec:results}. Spectral and pileup issues of these flares are discussed in Section \ref{sec:pileup}. Finally, discussion of these results is given in Section \ref{sec:discussion}.

\section{Event and Spike Selection}

\begin{figure}[]
\begin{center}
	\includegraphics[width=1.0\linewidth]{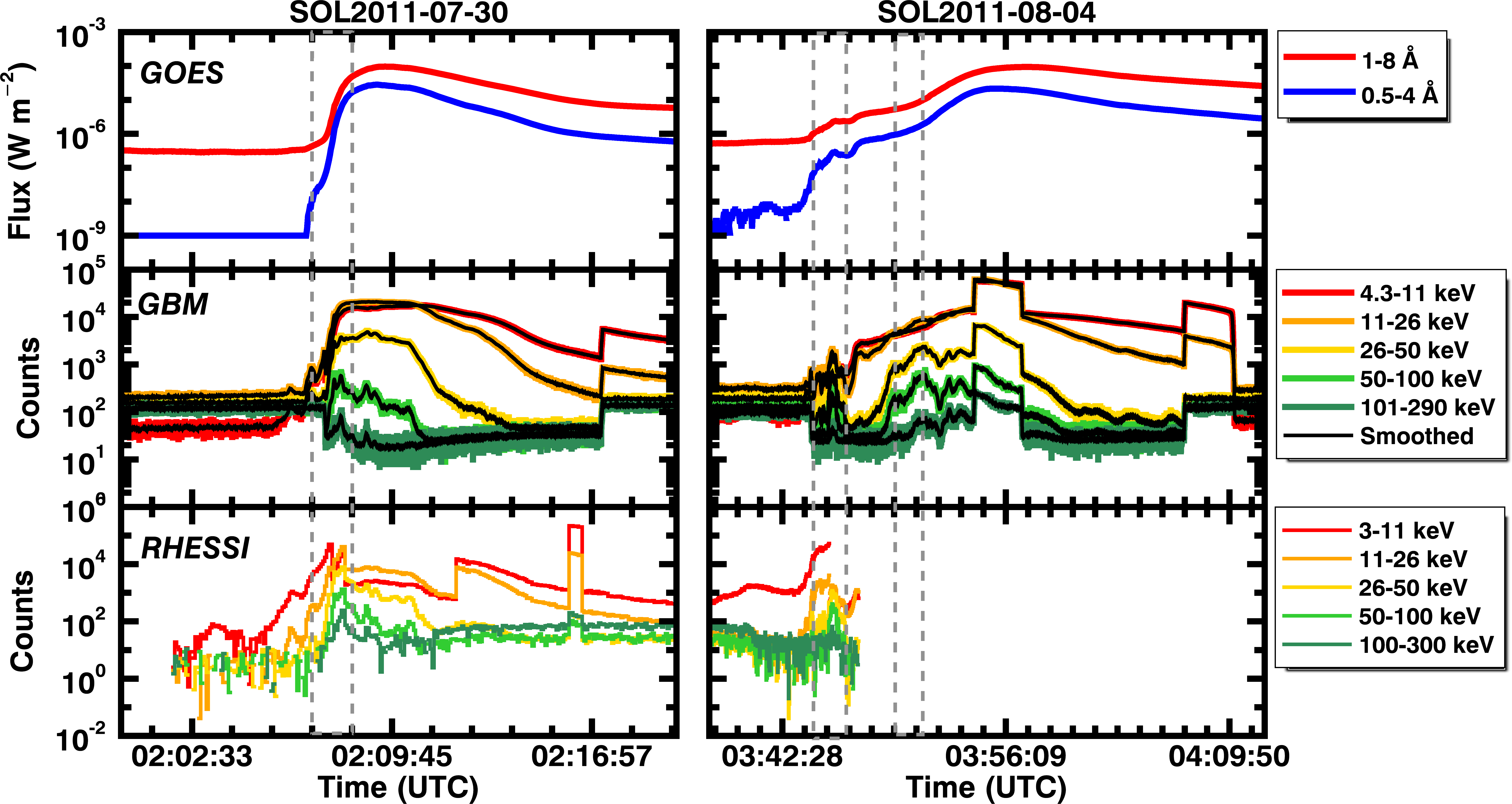}
\caption{ X-ray time profiles for the SOL2011-07-30 and SOL2011-08-04 M9.3 GOES-class solar flares. \textbf{Top:} \textit{GOES} SXR light curves. \textbf{Middle:} ``CTIME" \textit{Fermi} GBM HXR light curves. These light curves are a summation of the NaI(Tl) detectors, n1, n3, and n5, as these were the most sunward. The black curves represent a 4-second smoothing of the time profiles and are used in identifying subsecond spikes, as described in Section \ref{sec:identify}. \textbf{Bottom:} \textit{RHESSI} hard X-ray time profiles. \textit{RHESSI} data were used primarily for spectroscopy and comparison. The SOL2011-08-04 flare was only partially observed by \textit{RHESSI}. The gray lines indicate times when fast time spikes were identified. These intervals can be seen in Figure \ref{fig:intervals} }.
\label{fig:flare}
\end{center}
\end{figure}

\subsection{Flare Selection}\label{sec:selection}
\begin{figure}[]
\begin{center}
	\includegraphics[width=0.45\linewidth]{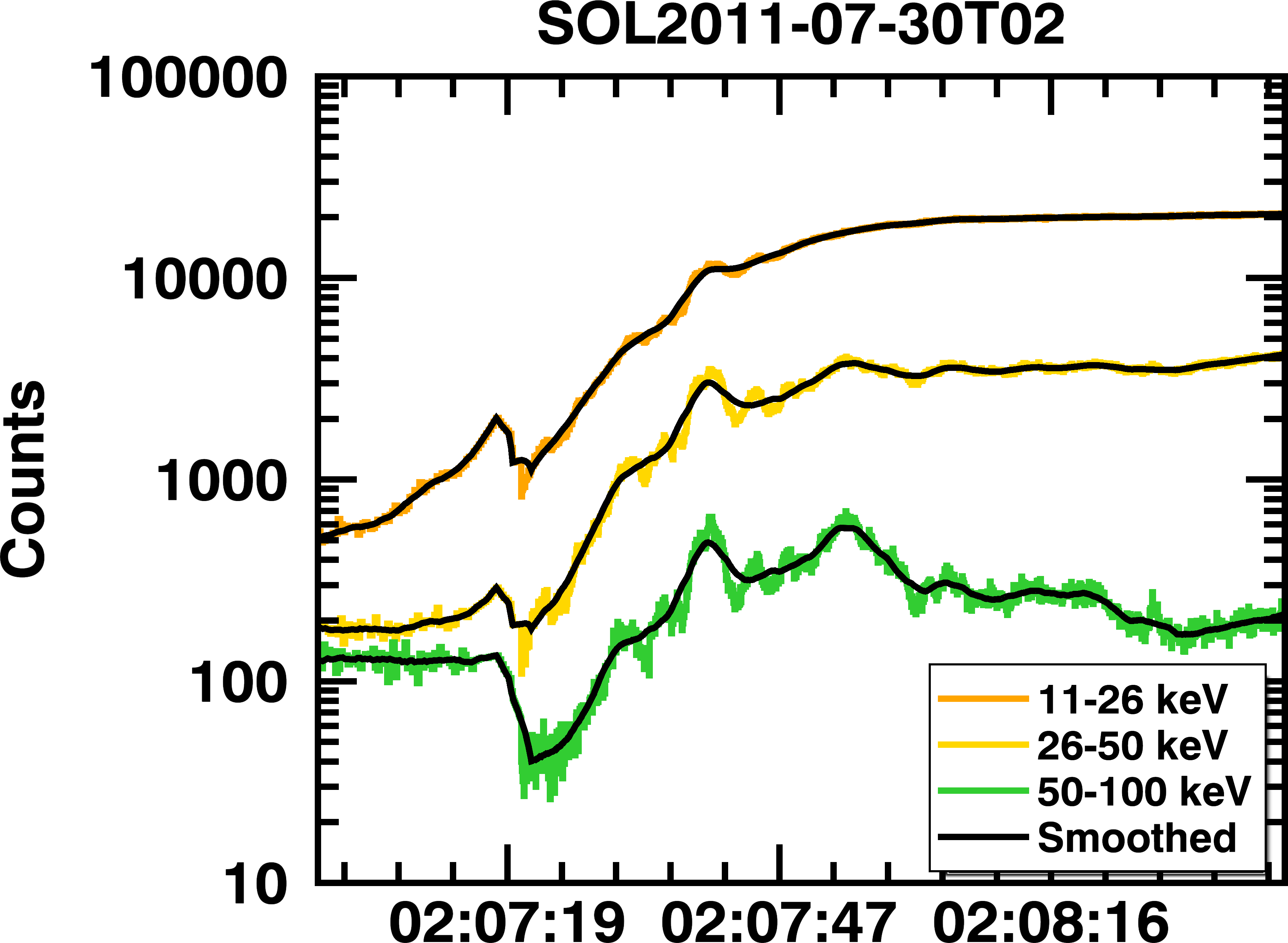}
	\includegraphics[width=0.45\linewidth]{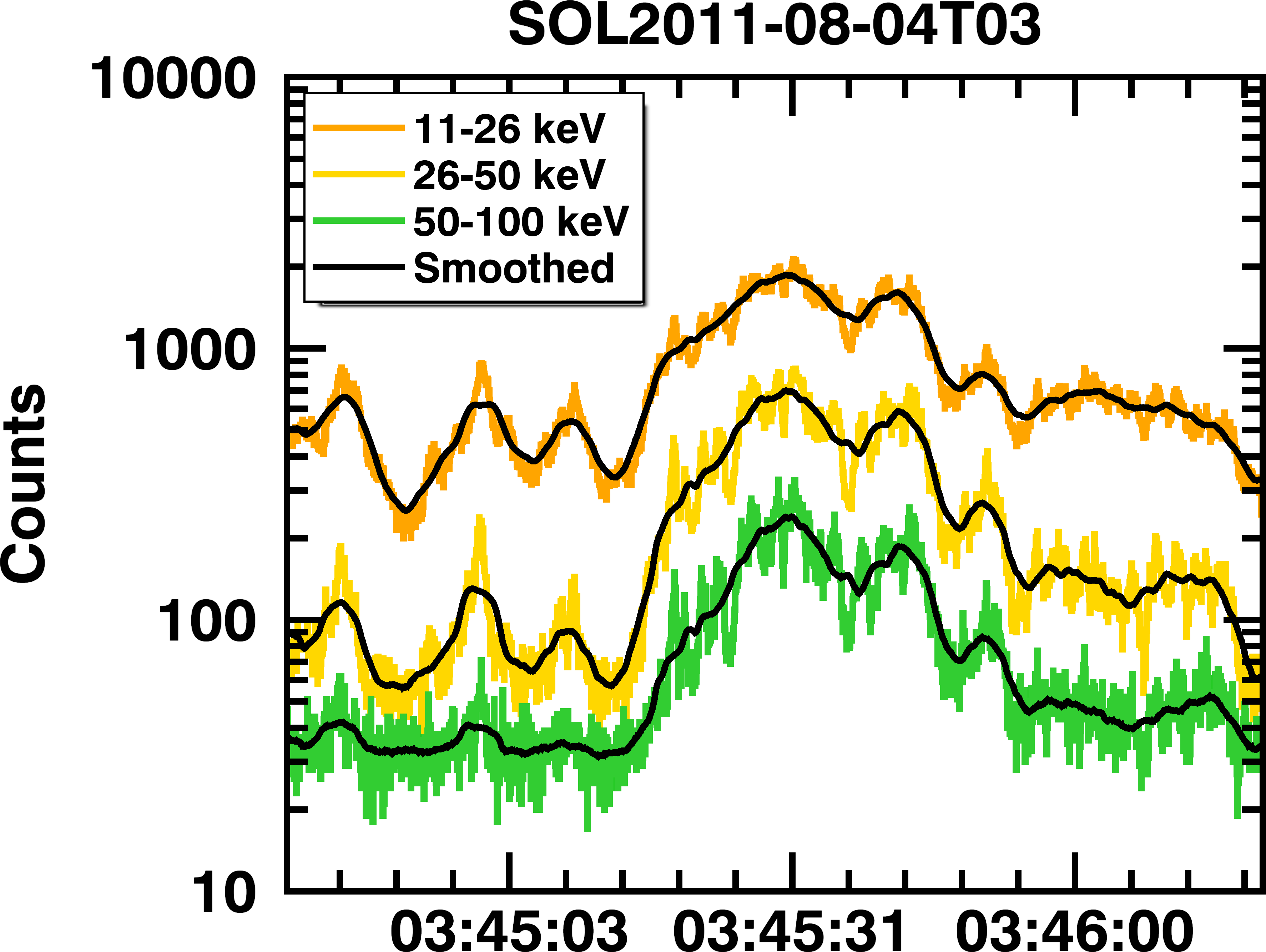}
	\includegraphics[width=0.45\linewidth]{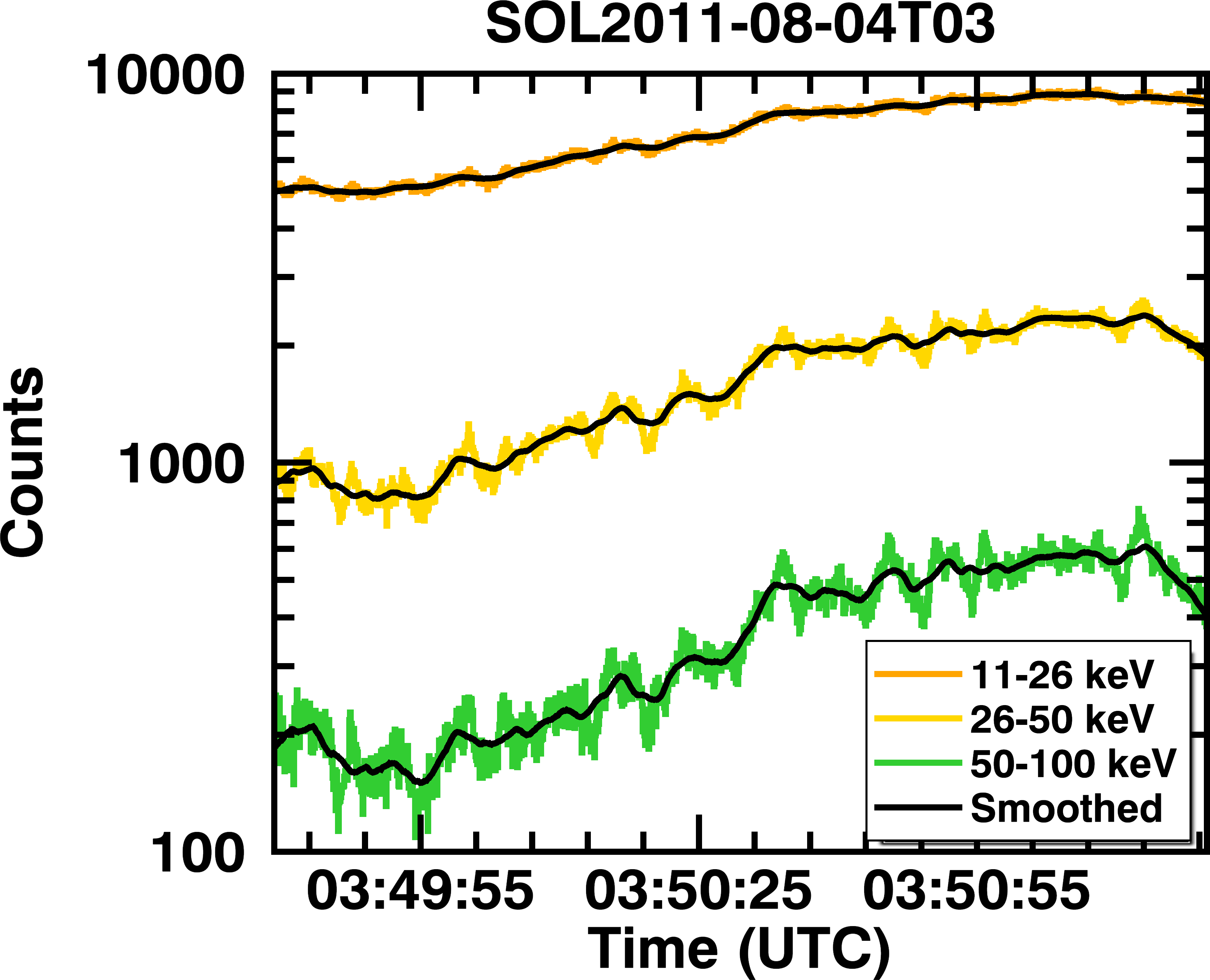}
\caption{ Highlighted portions of the flare time profiles where subsecond spikes were identified. The black curves are the 4-second smoothing applied to approximate the slowly varying component of the flare. The residuals for the entire top right plot can be seen in Figure \ref{fig:spikes}.}
\label{fig:intervals} 
\end{center}
\end{figure}

Our flare selection began by using the Interactive Multi-Instrument Database of Solar Flares\footnote{https://helioportal.nas.nasa.gov/} to find flares that were observed by both \textit{RHESSI} and \textit{Fermi}. Furthermore, we wanted to ensure good photon counting statistics without significant pulse-pileup and so flares of \textit{GOES}-class between M1 and X1 were selected. The 2 chosen flares are a pair of M9.3-class flares that occurred on July 30 and August 04, 2011. They both originated from Active Region NOAA 1261 and lasted for 26m 28s and 26m 17s, respectively, according to the \textit{Fermi} flare catalog. Figure \ref{fig:flare} shows the \textit{GOES}, \textit{Fermi}, and \textit{RHESSI} time profiles for these flares and Figure \ref{fig:intervals} shows a zoomed-in view of the times of interest.

\subsection{Subsecond Spike Identification}\label{sec:identify}
\begin{figure}[]
\begin{center}
	\includegraphics[width=0.45\linewidth]{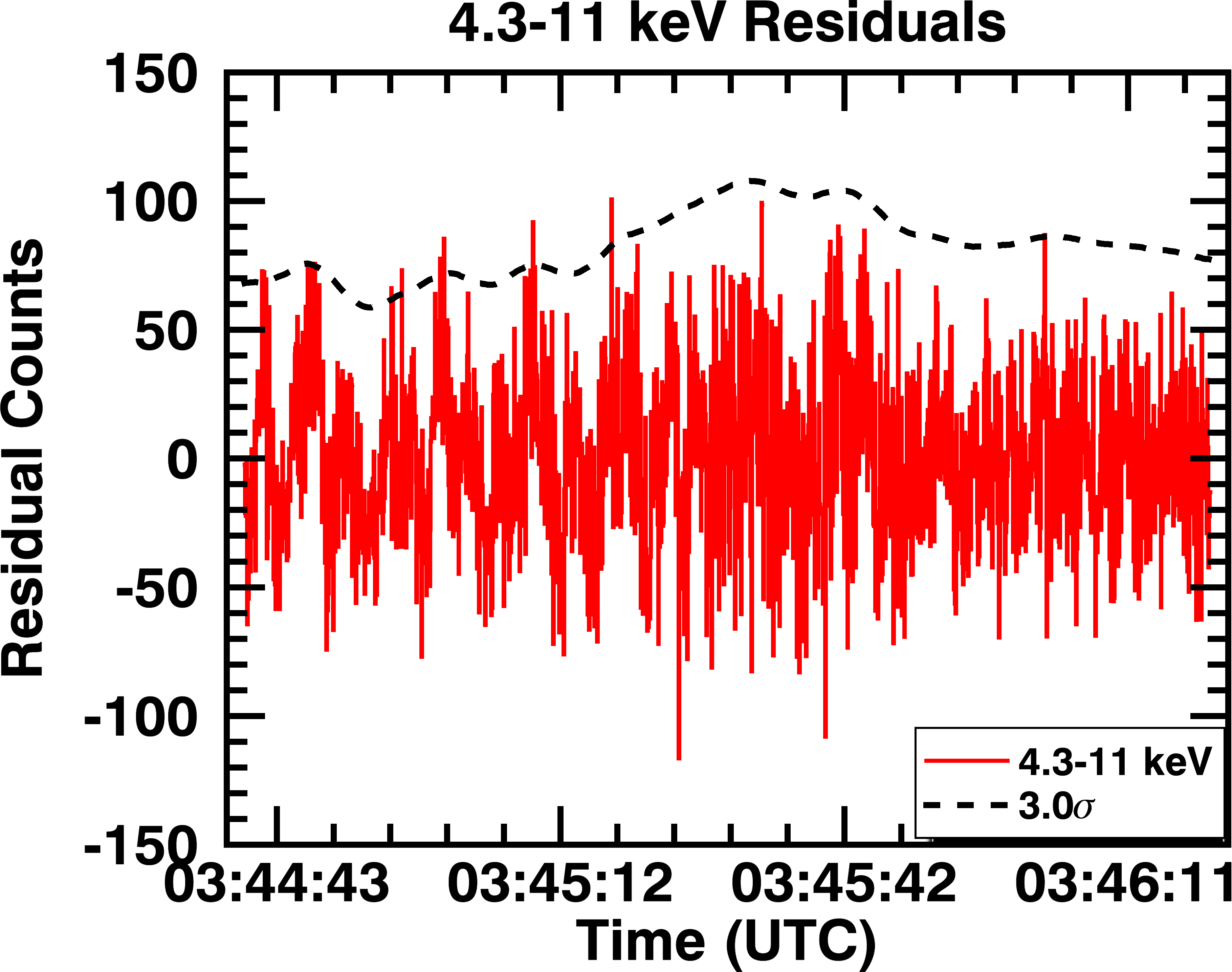}
	\includegraphics[width=0.45\linewidth]{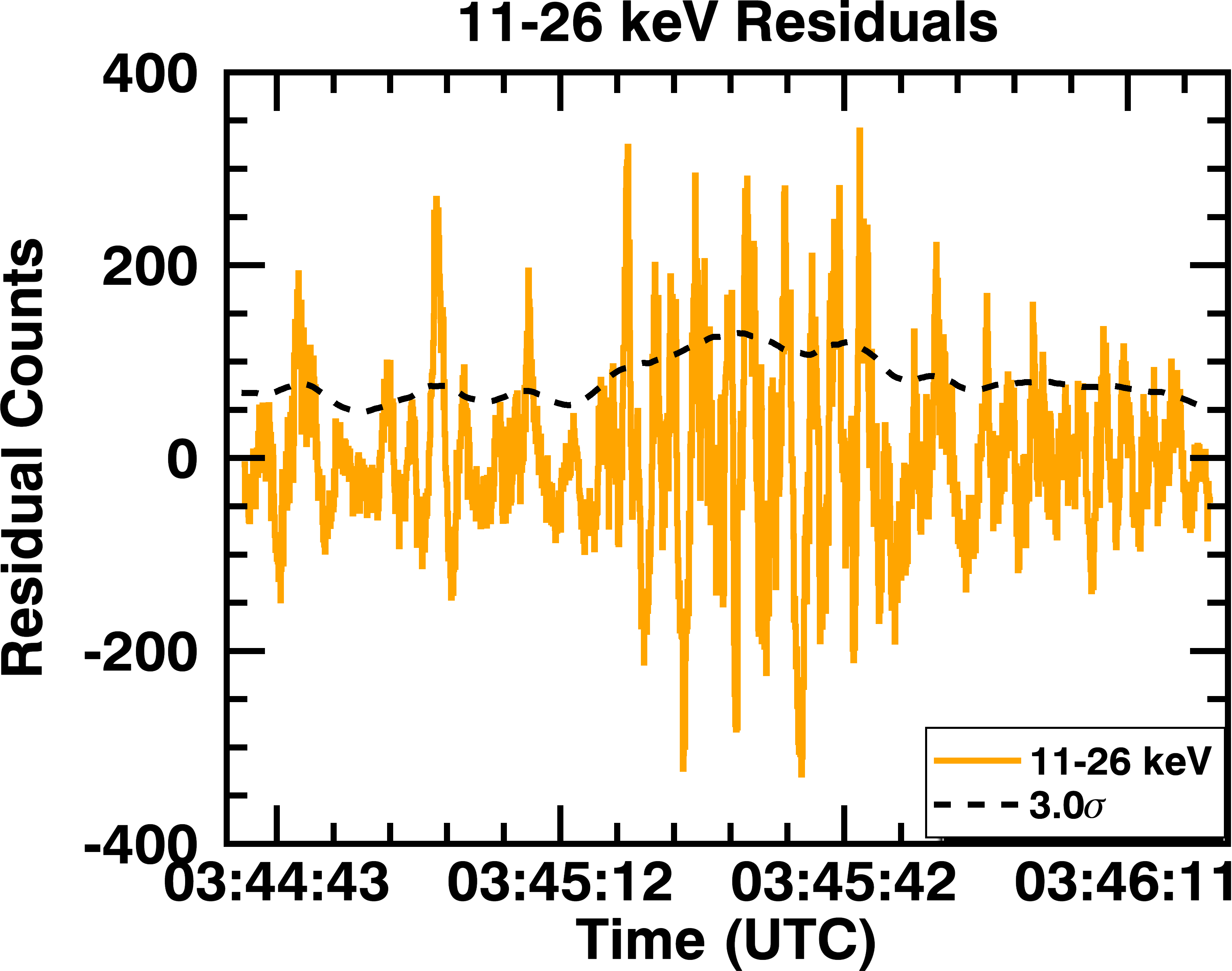}
	\includegraphics[width=0.45\linewidth]{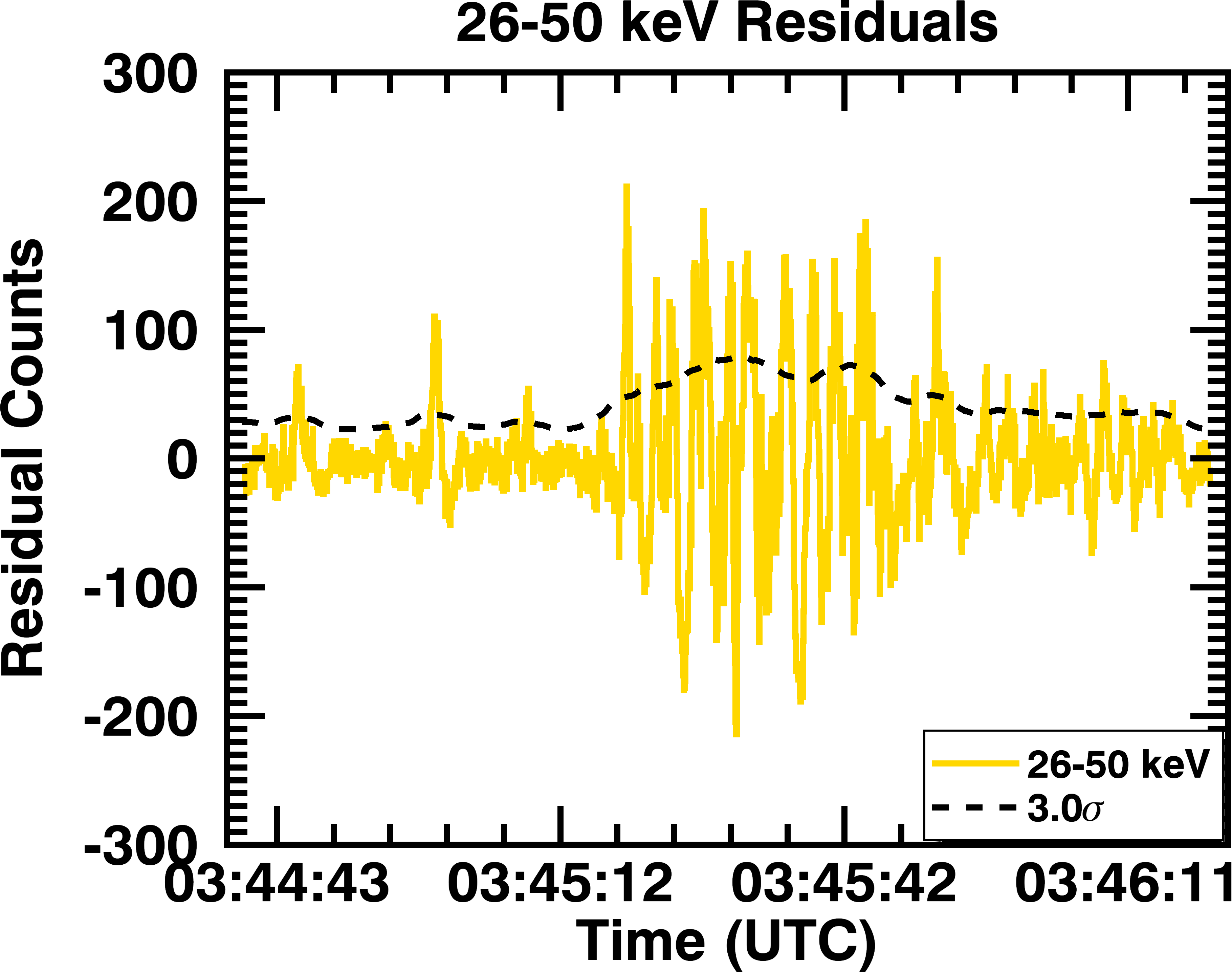}
	\includegraphics[width=0.45\linewidth]{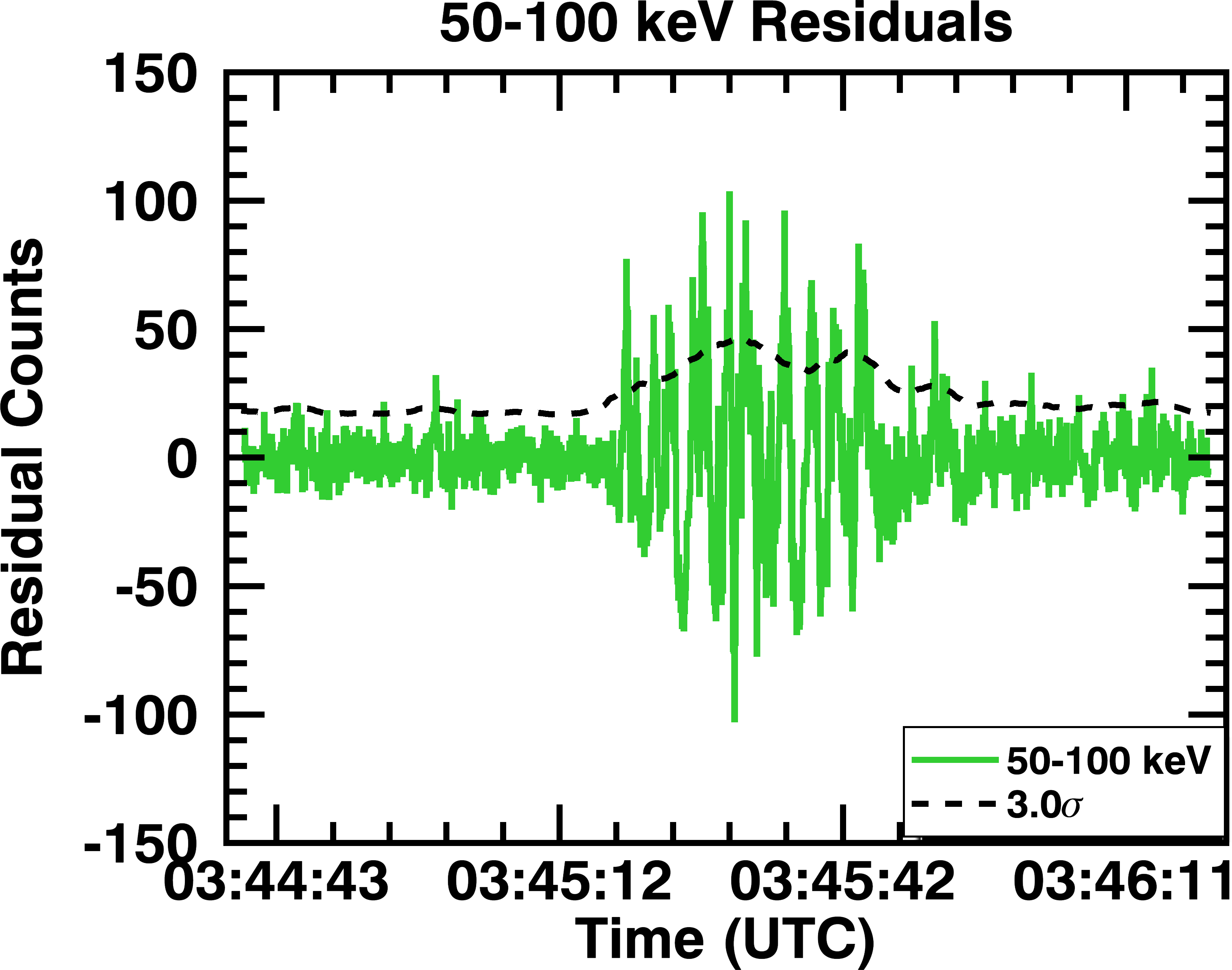}
\caption{ Subsecond residuals from the early interval of the SOL2011-08-04 flare. These short timescale fluctuations were identified after the black curves in Figures \ref{fig:flare} and \ref{fig:intervals} were subtracted from the raw counts. The black lines in these figures indicate the $3\sigma$ deviation from the smoothed curve. Spikes were identified as any fluctuation surpassing this threshold for a minimum of 3 time bins in a minimum of 2 adjacent energy channels. The 4.3-11 keV energy band failed to have any statistically significant spikes, presumably because it is dominated by thermal flux.}
\label{fig:spikes} 
\end{center}
\end{figure}

The \textit{Fermi} GBM ``CTIME" data product consists of 8 energy channels spanning 8 to 2000 keV with a roughly logarithmic binning. This format also has a 0.25 second time bin, unless overall flux rates exceed an internal threshold, in which case the time binwidth decreases to 0.0625 seconds. Identifying spikes in these light curves is based on the methods described in \cite{qiu}. The basis of this analysis is the assumption that the X-ray signal, S, can be represented as $S = s_{slow} + s_{fast} + s_{noise}$. $s_{slow}$ represents processes that occur on the order of several seconds or longer, $s_{fast}$ represents processes occurring on timescales of a few seconds or shorter, and $s_{noise}$ represents noise contributions due to Poisson statistics, background, and instrumental effects. To model $s_{slow}$, we apply a 4-second moving boxcar average to each energy channel. The black curves in Figure \ref{fig:flare} show the result of this smoothing. A standard deviation, $\sigma$, is then calculated from this smoothed curve as $\sigma = \sqrt{s_{slow}}$. Residuals are found by subtracting the smoothed curve from the raw counts. The next step is to identify signals that greatly exceed the statistical noise. To this end, any fluctuation exceeding a $3\sigma$ threshold is examined. If an apparent spike in the lightcurve persists above this threshold for a minimum of 3 consecutive time bins in a minimum of 2 energy channels, it is taken to be a real variation. Figure \ref{fig:spikes} shows an example of these subsecond variations across different energy bins. As we are examining short timescale signals from solar flare X-ray flux for the first time using \textit{Fermi} GBM data (a non-solar instrument), stringent selection criteria were used. For the time profiles examined, these selection criteria yield an expectation value of $1.5 \times 10^{-9}$ and $2.7 \times 10^{-9}$ false positives in the SOL2011-07-30 and SOL2011-08-04 flares, respectively. Relaxing these conservative criteria in future studies would allow for the identification of even shorter timescale spikes. This process identified spikes in the 11-26 keV, 26-50 keV, and 50-100 keV energy bands in both flares. Spikes can also be observed in the 4-11 keV band (see Figure \ref{fig:spikes}), but these were not statistically significant. It also revealed that these subsecond spikes occurred in intervals of approximately 60-100 seconds. 1 such interval was identified in the SOL2011-07-30 flare from 02:07:20 - 02:08:20 UTC. 2 were identified in the SOL2011-08-04 flare from 03:45:00 - 03:46:00 and 03:49:40 - 03:51:20 UTC.

\subsection{Spike Shape Fitting}\label{sec:gmm}

\begin{figure}[]
\begin{center}
	\includegraphics[width=0.45\linewidth]{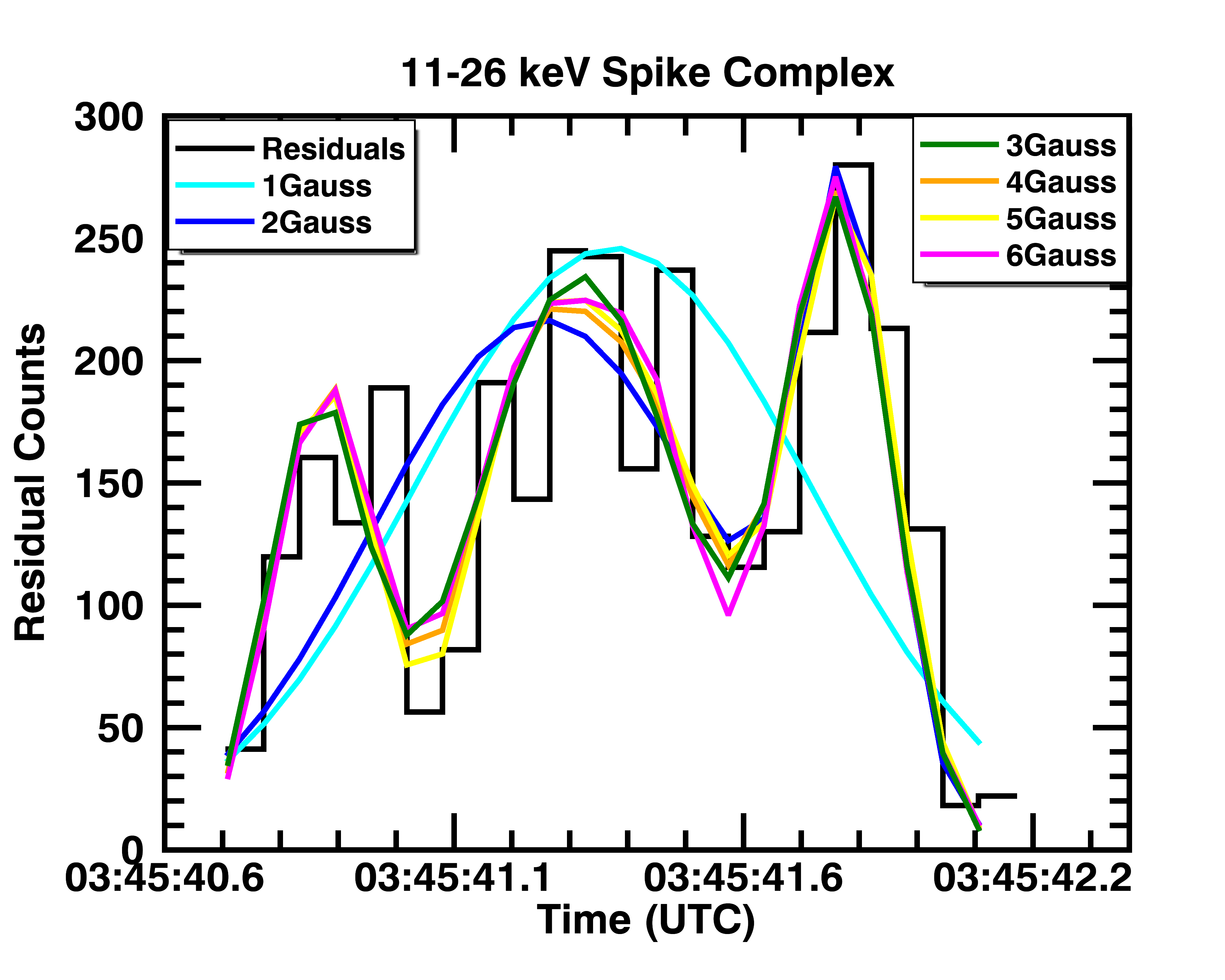}
	\includegraphics[width=0.45\linewidth]{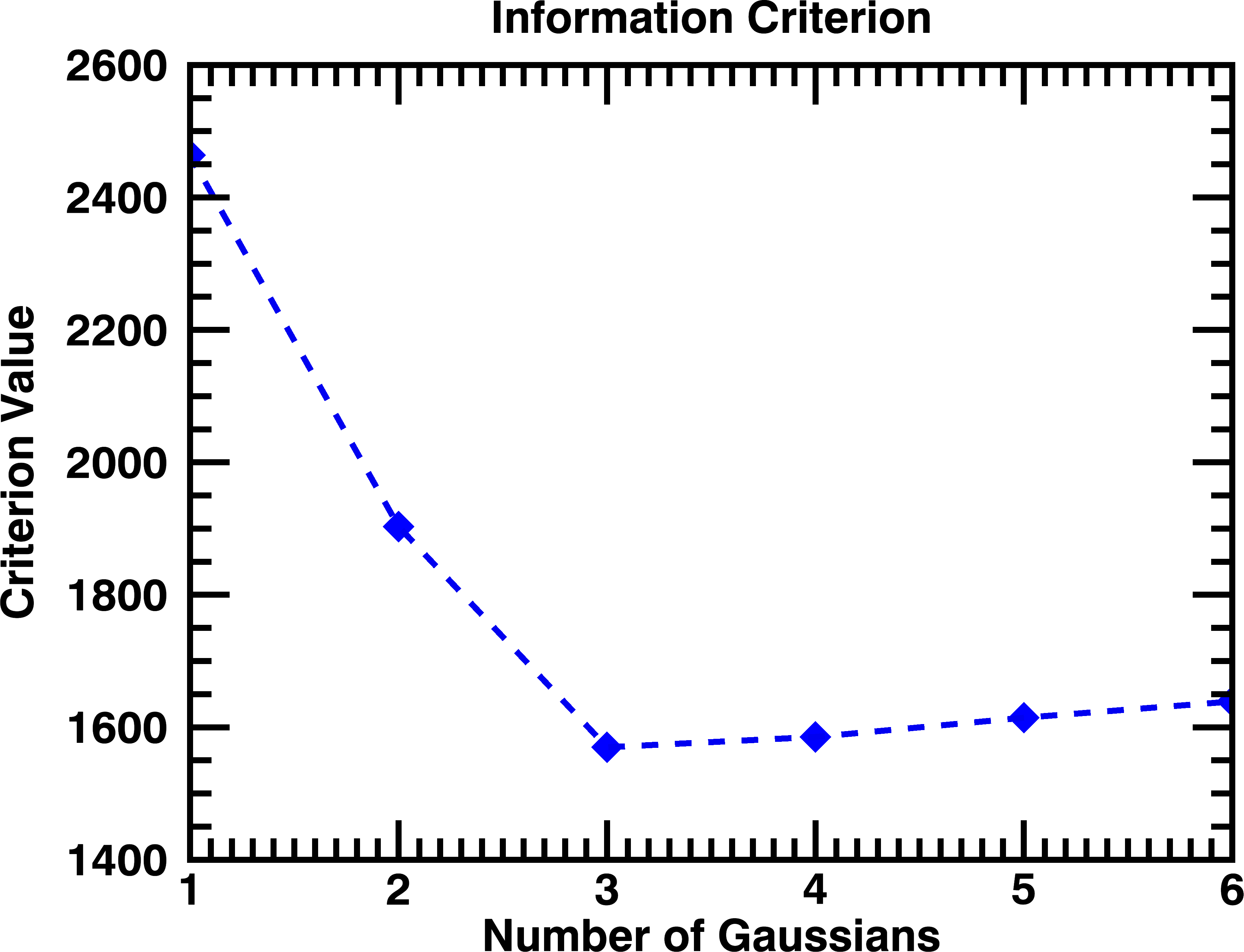}
\caption{An example of the Gaussian Mixture Model (GMM) expectation maximization approach. This section of data, taken from the SOL2011-08-04 flare, demonstrates that rapid subsecond spikes can combine to create a single spike complex that will be flagged as 1 individual spike by our criteria. These complexes are analyzed with a GMM with the optimal number of spikes being determined by the Bayesian Information Criterion (BIC). For this spike complex, the best-fit number of Gaussians is 3. This information is then passed to the curve fitting routine being applied to the raw counts, as shown in Figure \ref{fig:rawcounts}.}
\label{fig:gmm_example}
\end{center}
\end{figure}

With strong subsecond spike candidates identified, it becomes necessary to determine their physical characteristics, the most important being their peak times and durations. Fitting the spikes is a difficult endeavor owing to the fact that they rest upon a slowly-varying background that is not well modeled from first principles in relatively short (10-30 second) timeframes. Additionally, there is an issue of spikes occurring in quick repetition and forming a large spike complex that was detected in the previous step as a single spike. The method we have used to handle these issues is to use an expectation maximization (EM) approach to fit a Gaussian Mixture Model (GMM) to each detected spike in the residual count space. As the number of spikes is not known a priori, the GMM is represented as the sum of 1 to 6 Gaussian functions. With maximum likelihoods calculated for a fit to each number of Gaussians, a Bayesian Information Criterion, $BIC = \textrm{ln}(n)k - 2\textrm{ln}(\hat{L})$, was applied. $n$ is the number of data points being fitted, $k$ is the number of parameters in the fitted model, and $\hat{L}$ is the maximum likelihood of each fit. The fitting bounds were determined by expanding the bounds about each identified spike until negative residuals were encountered. Figure \ref{fig:gmm_example} shows an example of a spike complex best fit by 3 Gaussians. GMM fits for 1 through 6 Gaussians are displayed with their corresponding BIC. The minimum of the BIC curve indicates the best estimate for the number of Gaussians. This is due to the fact that the BIC penalizes models for free parameters and rewards them for better fits. The total number of spikes found via the GMM algorithm in the SOL2011-07-30 flare are 15, 22, and 7 for the 11-26, 26-50, and 50-100 keV energy bands, respectively. For the SOL2011-08-04 flare they were 52, 69, and 43. 

\begin{figure}[]
\begin{center}
	\includegraphics[width=0.5\linewidth]{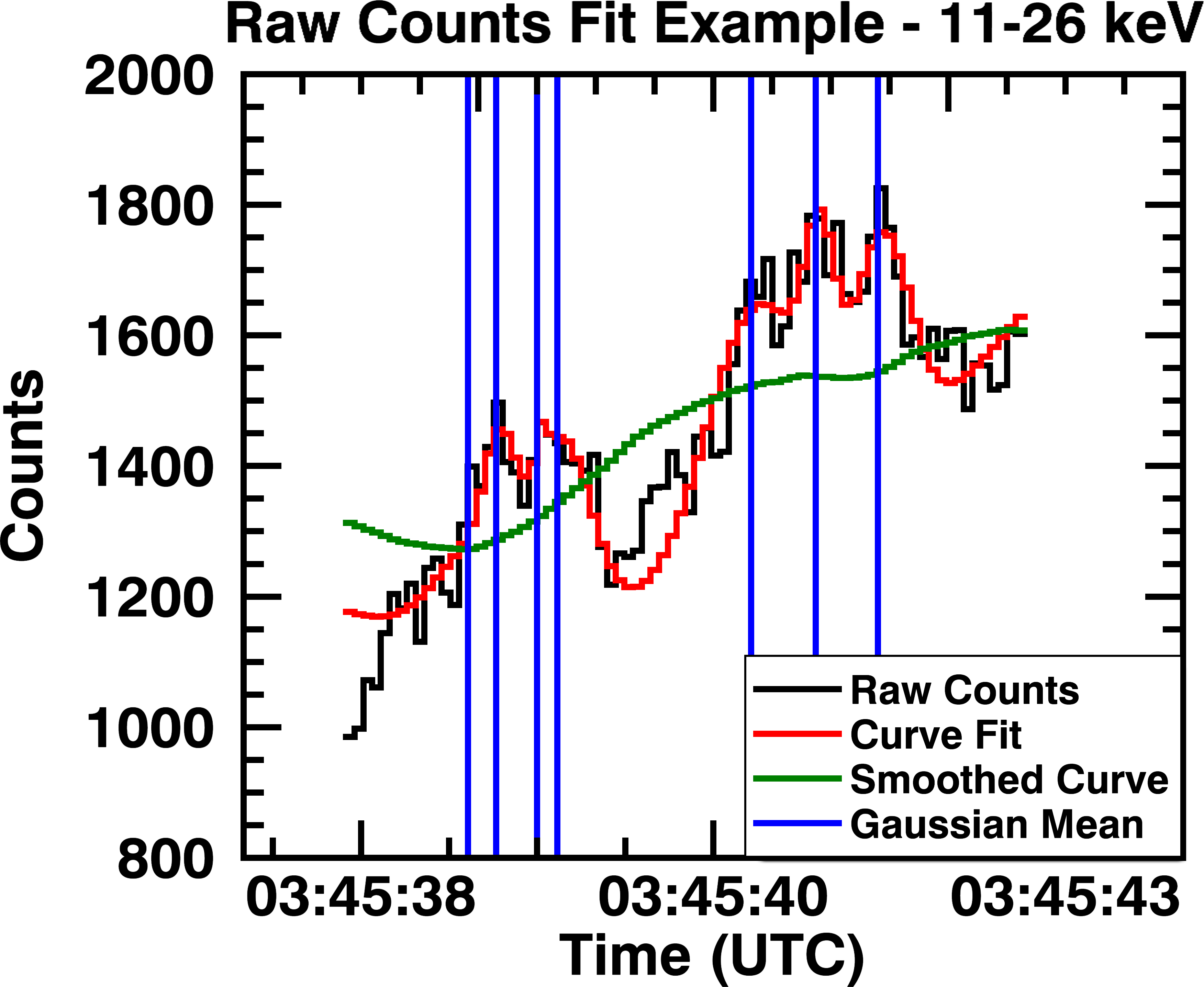}
\caption{ An example of multi-Gaussian fitting with a quadratic background. The green curve represents the 4-second smoothing described in Section \ref{sec:identify}. The red line represents the fitted curve. The fit does not fit the background perfectly since the backgrond is not purely quadratic, but it provides a significantly better baseline than the 4-second smoothed curve for fitting the spikes which occur at the vertical blue lines.}
\label{fig:rawcounts}
\end{center}
\end{figure}

The means and number of Gaussians were then input into a fitting algorithm that progressed through the raw counts (not the residuals) at nominally 8-second intervals, beginning at the flare's start time according to the \textit{Fermi} flare catalog. Any fitting interval that had a spike within 1.5 seconds of a boundary was expanded to include the entirety of the spike. The routine fit the signal with a function consisting of a quadratic background (to account for slow variations) combined with multiple Gaussians using a nonlinear least squares method. The form of this function is $S(t) = at^{2} + bt + c + \sum_{i}^{N} \alpha_{i} exp(-\frac{(\mu_{i} - t)^{2}}{\sigma_{i}^{2}})$. $a$, $b$, and $c$ are the parameters for the quadratic background while $\alpha_{i}$, $\mu_{i}$, $\sigma_{i}$ are the amplitudes, means, and standard deviations for the Gaussians, respectively, with the total number, $N$, being determined by the EM algorithm. Figure \ref{fig:rawcounts} shows an example of this fitting routine.

We also examined the relative spike intensity. By dividing each spike's amplitude by the quadratic background, we calculated the peak intensity of each spike in each energy band. Table \ref{tab:intensity} shows the results of these calculations. In general, the relative intensity of a spike increases with energy. This is in line with our expectation that these spikes are a purely nonthermal phenomenon, and nonthermal X-ray flux becomes more dominant over thermal flux at higher energies.

\begin{table}[]
\centering
\caption{Measurements of spike intensity relative to the slowly varying background. Spikes identified in higher energy bands tend to be more intense, likely due to the lack of thermal emission at those energies. }
\begin{tabular}{lllll}
\hline
\multicolumn{5}{c}{\textbf{Spike Intensity (Fitted Spike Amplitude/Background)}}                                                                                                                                              \\ \hline
\multicolumn{1}{c}{\textbf{Flare}} & \multicolumn{1}{c}{\textbf{Measurement}} & \multicolumn{1}{c}{\textbf{11-26 keV}} & \multicolumn{1}{c}{\textbf{26-50 keV}} & \multicolumn{1}{c}{\textbf{50-100 keV}} \\ \hline \hline
\multirow{5}{*}{SOL2011-07-30}           & \# of Spikes                             & 15                                     & 22                                     & 7                                       \\
                                   & Minimum                                  & 0.02                                   & 0.01                                   & 0.22                                    \\
                                   & Median                                   & 0.06                                   & 0.10                                   & 0.28                                    \\
                                   & Mean                                     & 0.09                                   & 0.10                                   & 0.28                                    \\
                                   & Maximum                                  & 0.27                                   & 0.32                                   & 0.39                                    \\ \hline
\multirow{5}{*}{SOL2011-08-04}         & \# of Spikes                             & 52                                     & 69                                     & 43                                      \\
                                   & Minimum                                  & 0.01                                   & 0.03                                   & 0.05                                    \\
                                   & Median                                   & 0.16                                   & 0.22                                   & 0.23                                    \\
                                   & Mean                                     & 0.20                                   & 0.31                                   & 0.27                                    \\
                                   & Maximum                                  & 0.94                                   & 1.54                                   & 0.39                                    \\ \hline
\label{tab:intensity}
\end{tabular}
\end{table}

\section{Spike Analysis and Results}\label{sec:results}
With the spikes positively identified by our selection criteria and then fit using a GMM approach, we have a list of spikes and their properties. We then analyze these spike properties in a variety of ways, namely their duration distributions, lag time between energy bands and quasi-periodicities.

\subsection{Spike Durations}
\begin{figure}[]
\begin{center}
	\includegraphics[width=0.45\linewidth]{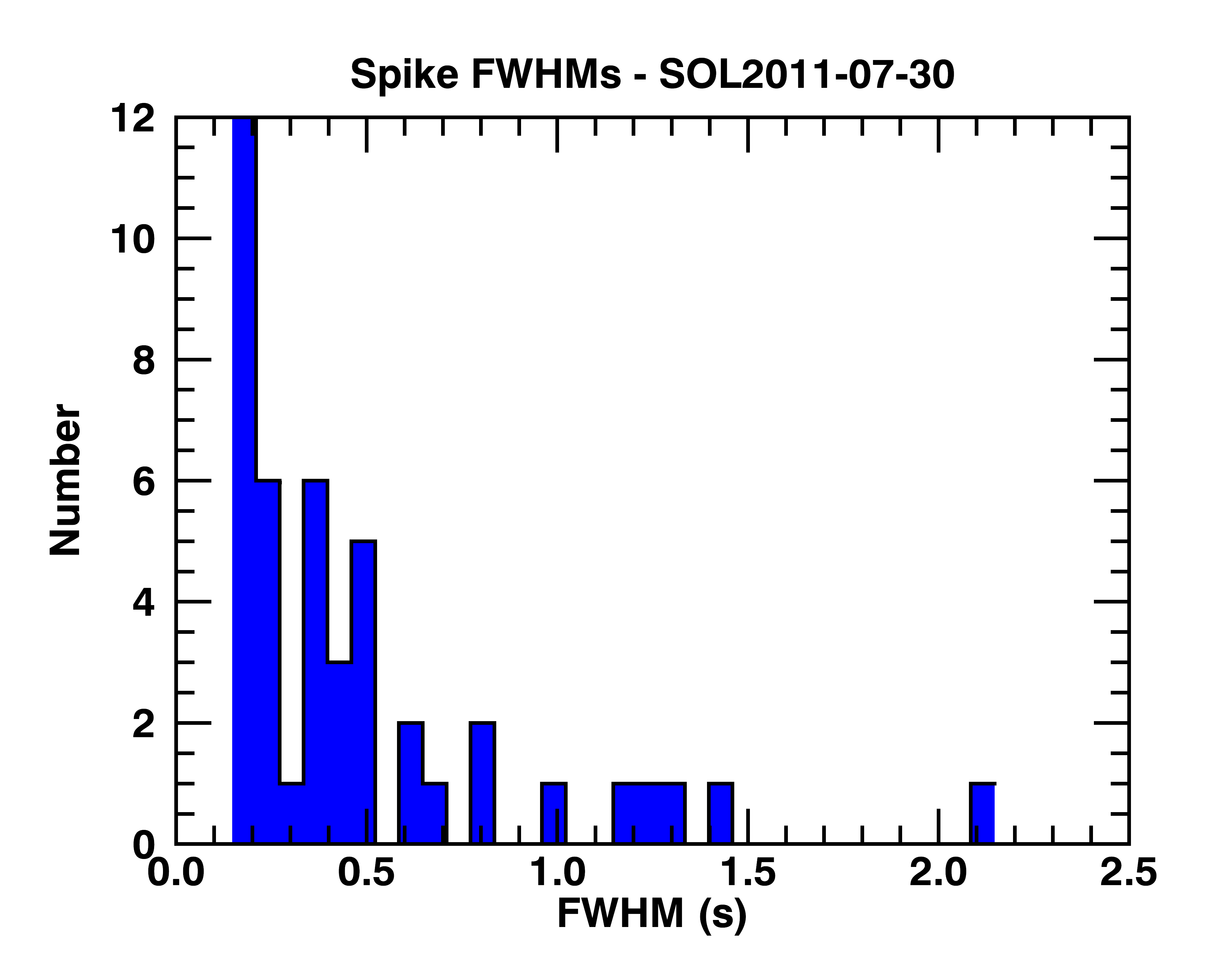}
	\includegraphics[width=0.45\linewidth]{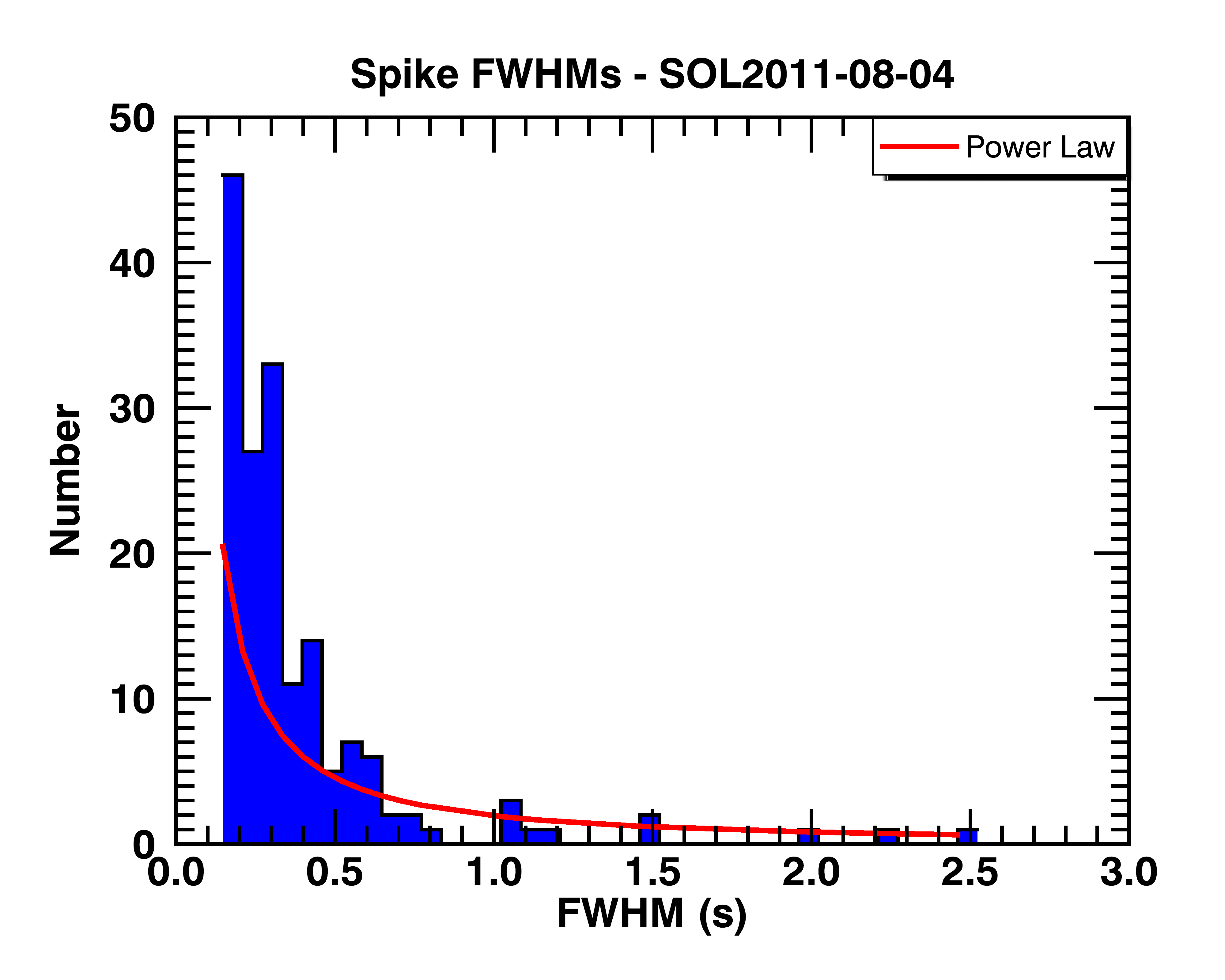}
\caption{ The distribution of spike full width half-maximums (FWHMs). The SOL2011-08-04 flare featured a far larger number of short timescale spikes than the SOL2011-07-30 flare. The mean and median of the spike duration distributions are 0.49 and 0.36 seconds for the SOL2011-07-30 flare and 0.38 and 0.30 seconds for the SOL2011-08-04 flare. A simple power law was also fit to the SOL2011-08-04 flare distribution and found to have an index of $-1.2 \pm 0.3$.}
\label{fig:dist}
\end{center}
\end{figure}

A subsecond spike in bremsstrahlung X-ray flux must be created via a subsecond burst of energetic electrons \citep{brown}. Figure \ref{fig:dist} shows the distributions of spike durations (Gaussian FWHMs) found in the 2 flares. The SOL2011-07-30 flare had a spike mean of 0.49 seconds and a median of 0.36 seconds. The SOL2011-08-04 flare had a mean of 0.38 seconds and a median of 0.30 seconds. Spikes very likely exist at shorter durations, but are not found in our data becuase of the chosen binning of Fermi GBM data and our selection criteria, which require emission above a threshold in three consecutive time bins. A simple power law ($ y = Ax^{b}$) was fit to the SOL2011-08-04 spike duration distribution. An index of $b= -1.2 \pm 0.3$ was found for this fit.

\subsection{Spike Lag Time}
According to the standard model of flares, particles are accelerated near the reconnection region and then propagated to the footpoints, where they emit hard X-rays. This means the particle time of flight is included within the X-ray spike duration. Differences in the peak times for spikes across energy bands can be used to infer time of particle accleration and time of flight timescales. For instance, high energy radiation peaking first implies a direct and rapid acceleration mechanism in the corona with higher energy particles (with a short time of flight) reaching the photosphere faster. Low energy radiation peaking first may imply a particle trapping mechanism (such as that in a magnetic mirror) with low energy particles escaping first. No major lag may imply a mixture of these phenomenon, or places a limit on the relevant timescales as being shorter than the binning timescales.

\begin{figure*}[htp]
\begin{center}
	\includegraphics[width=1.0\linewidth]{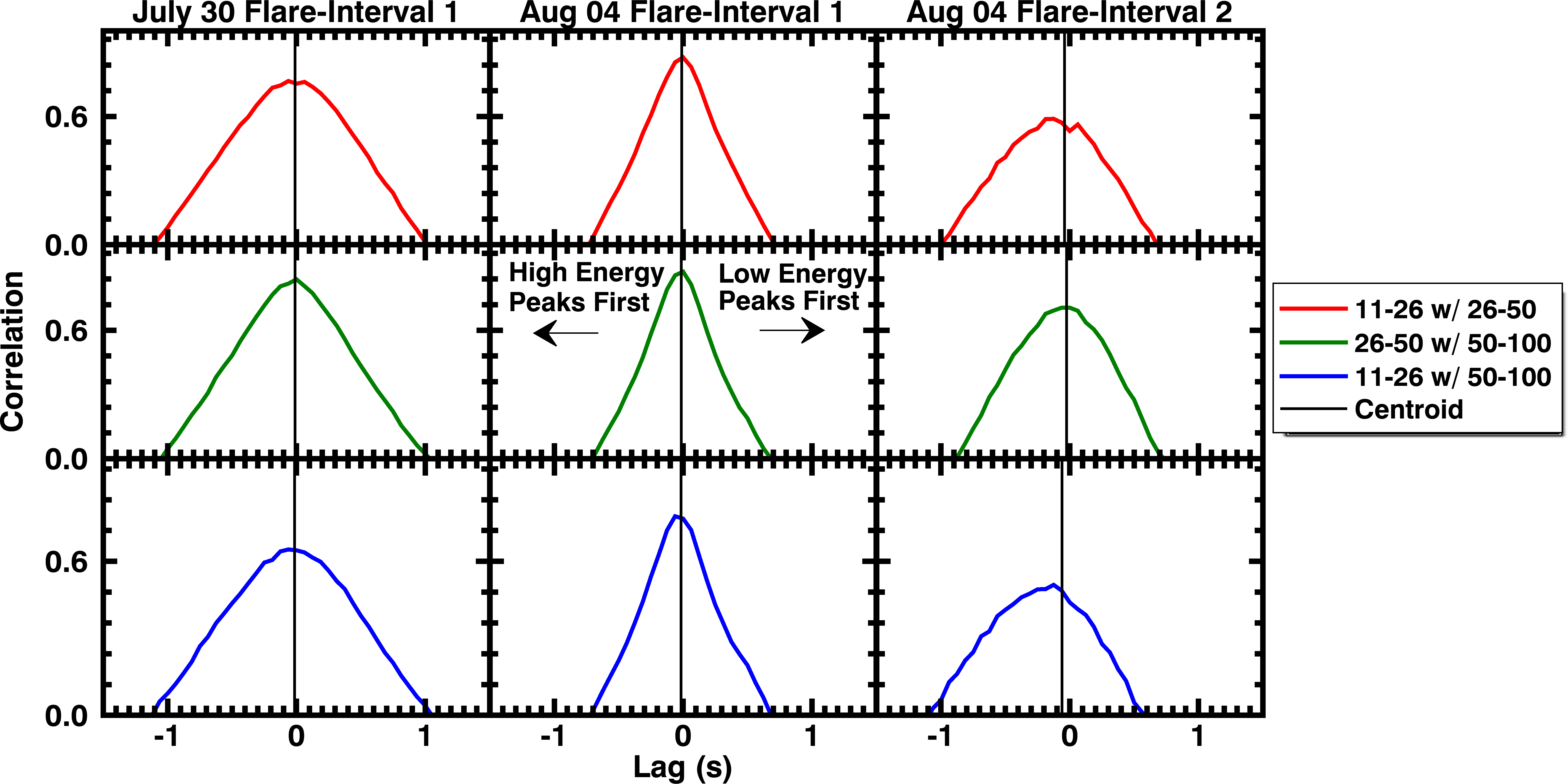}
\caption{ The cross correlation vs. lag for the 3 spiking intervals in the 2 flares. These plots measure the cross correlation in the residual count space of 2 energy bands vs. a lag between them. A negative lag corresponds to the high energy peaking first while a positive lag corresponds to the low energy peaking first. The black vertical lines indicate the centroid of the peaks.}
\label{fig:correlation}
\end{center}
\end{figure*}

A cross correlation analysis was performed on the residual counts of different energy bands for each the 3 intervals; 1 in the SOL2011-07-30 flare and 2 in the SOL2011-08-04 flare. The results of these correlations can be found in Table \ref{tab:corr}. A negative lag time indicates a peak correlation was found by moving the lower energy band earlier in time. Therefore, a negative lag time indicates that the higher energy component peaks first. Table \ref{tab:corr} shows that the spikes systematically peaked first in the higher energy bands. To estimate the uncertainty we calculated a centroid on the curves seen in Figure \ref{fig:correlation}. We varied the intervals about 0 seconds from $\pm$0.1875 seconds to $\pm$1 second and calculated the centroid for 13 intervals in total. We then calculated a standard deviation of these centroids and used it to indicate the uncertainty of the time lag. Pulse pileup can affect these results by reducing the observed difference in spike peaks across energy bands, leading these to be a lower limit on the lag. More discussion on the pulse pileup of the Fermi GBM data can be found in Section \ref{sec:pileup}.

\begin{table*}[htp]
\caption{The cross correlation vs. lag time was calculated for the 3 bursts of spikes. A centroid was used to determine the lag time with peak correlation. A negative time indicates higher energy spikes peaking first. }
\begin{tabular}{llll}
\hline
\multicolumn{4}{c}{\textbf{Lag Cross Correlation}}                                                                                                                 \\ \hline \hline
\multicolumn{1}{c}{\textbf{Flare}}        & \multicolumn{1}{c}{\textbf{Interval (UTC)}} & \textbf{Energy Bands (keV)} & \textbf{Peak Lag Time (s)} \\ \hline
\multirow{3}{*}{SOL2011-07-30}   & \multirow{3}{*}{2:07:20 - 2:08:20} & 11-26 w/ 26-50                               & -0.012 $\pm$ 0.010                            \\
                                 &                                    & 26-50 w/ 50-100                              & -0.011 $\pm$ 0.007                            \\
                                 &                                    & 11-26 w/ 50-100                              & -0.014 $\pm$ 0.011                           \\ \hline
\multirow{6}{*}{SOL2011-08-04} & \multirow{3}{*}{3:45:00 - 3:46:00} & 11-26 w/ 26-50                               & -0.009 $\pm$ 0.004                            \\
                                 &                                    & 26-50 w/ 50-100                              & -0.014 $\pm$ 0.005                            \\
                                 &                                    & 11-26 w/ 50-100                              & -0.016 $\pm$ 0.005                            \\ \cline{2-4} 
                                 & \multirow{3}{*}{3:49:40 - 3:51:20} & 11-26 w/ 26-50                               & -0.07 $\pm$ 0.03                              \\
                                 &                                    & 26-50 w/ 50-100                              & -0.03 $\pm$ 0.02                            \\
                                 &                                    & 11-26 w/ 50-100                              & -0.13 $\pm$ 0.08                              \\ \hline
\label{tab:corr}
\end{tabular}
\end{table*}

\subsection{Spike Periodicity}
\begin{figure*}[]
\begin{center}
	\includegraphics[width=0.45\linewidth]{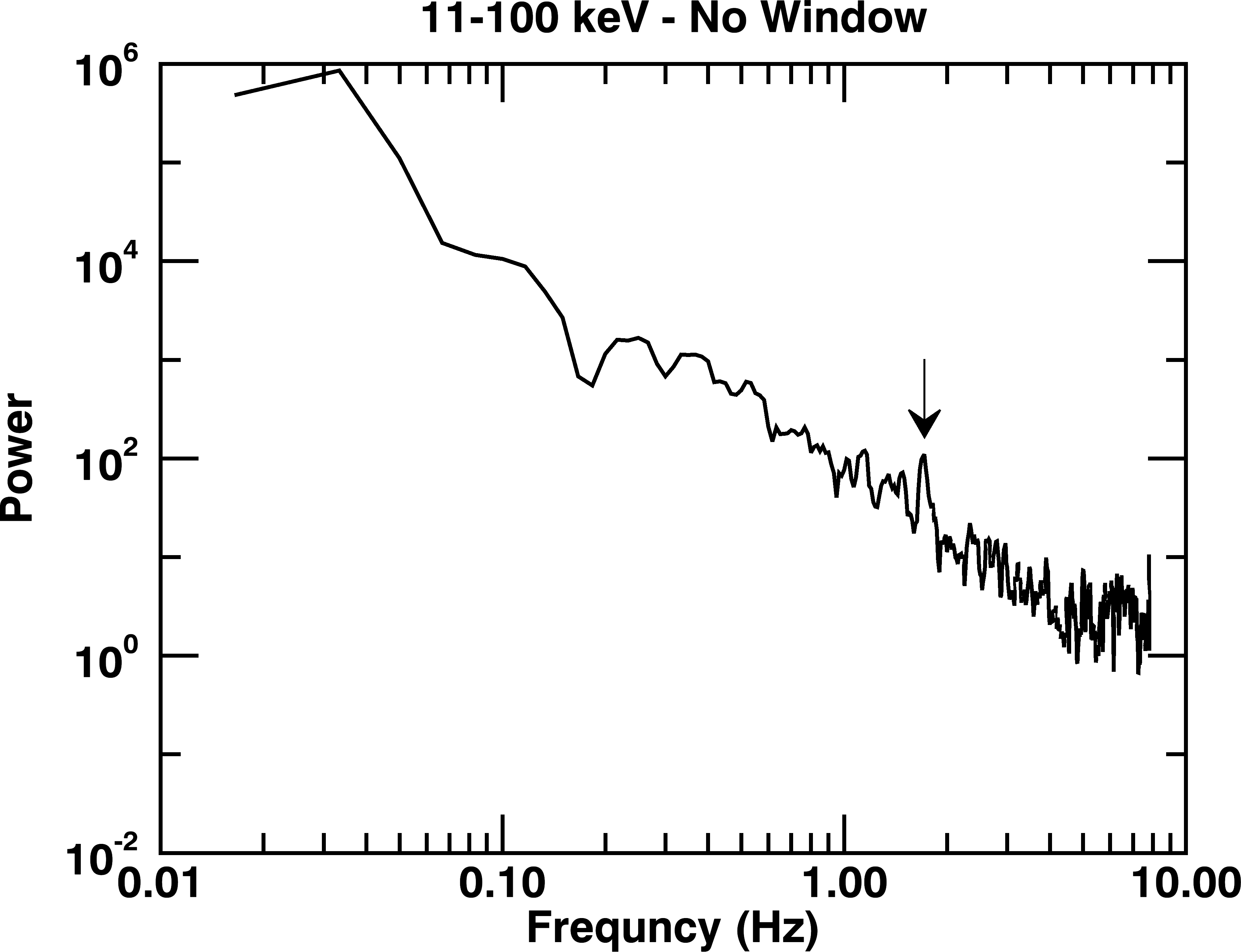}
	\includegraphics[width=0.45\linewidth]{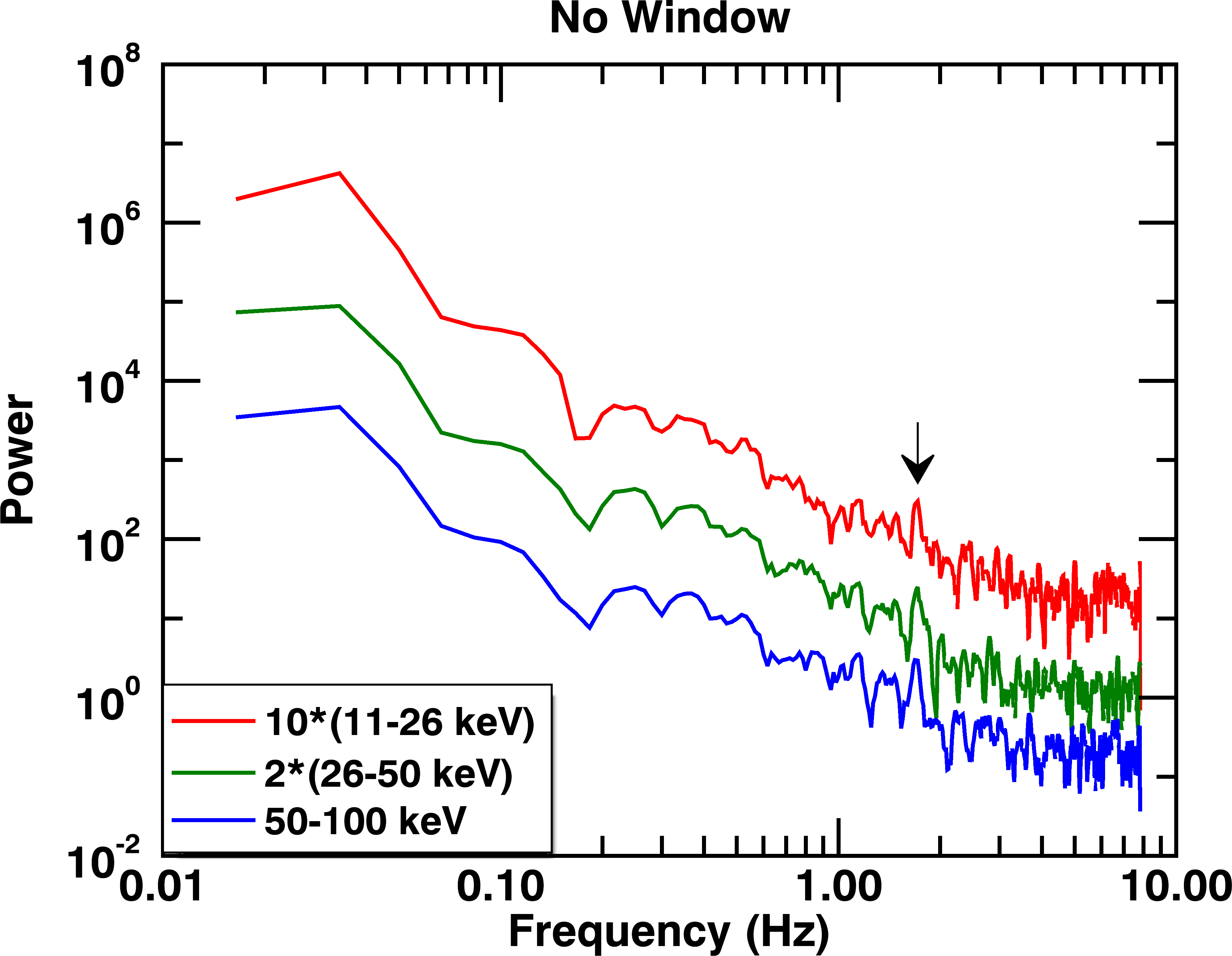}
	\includegraphics[width=0.45\linewidth]{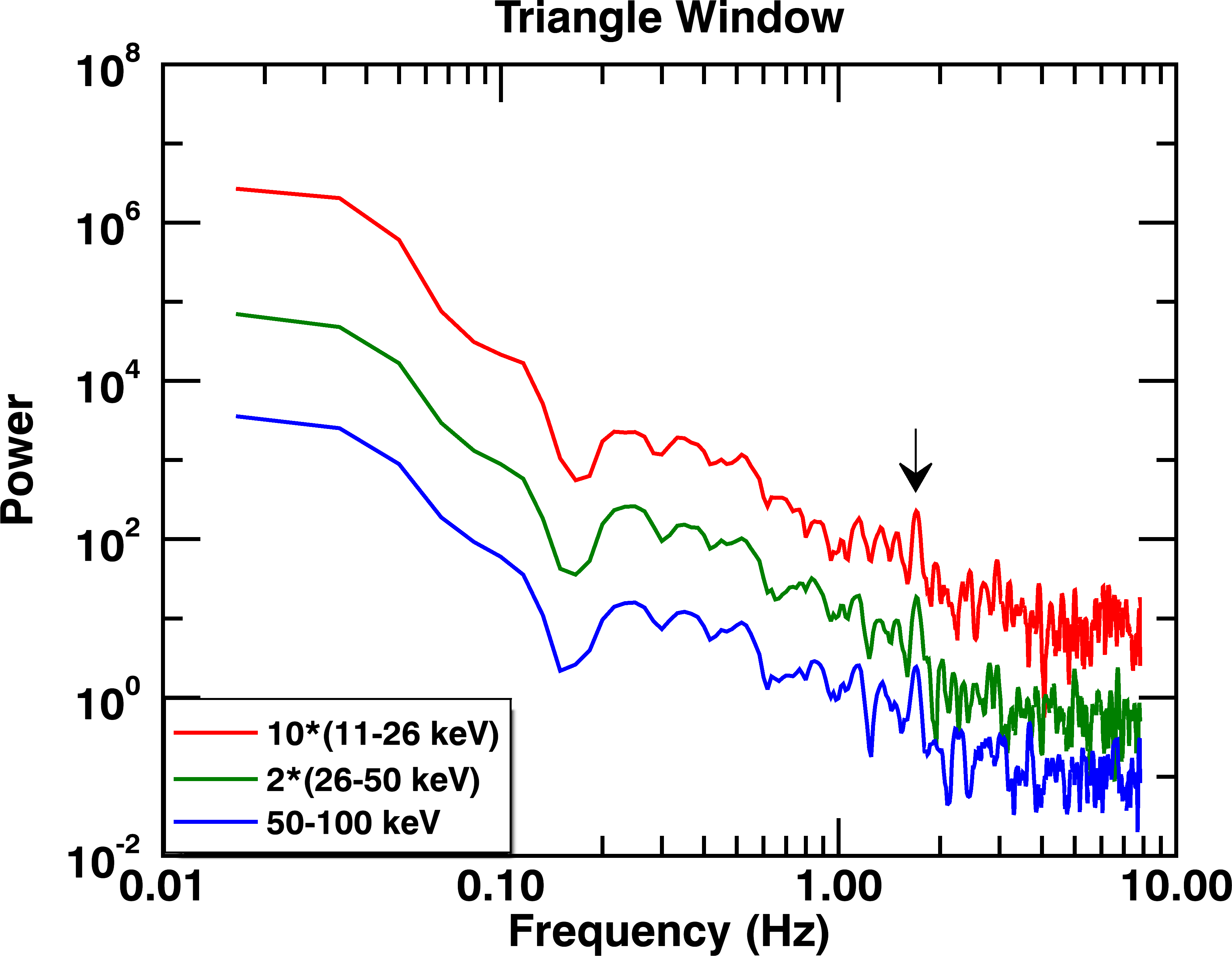}
	\includegraphics[width=0.45\linewidth]{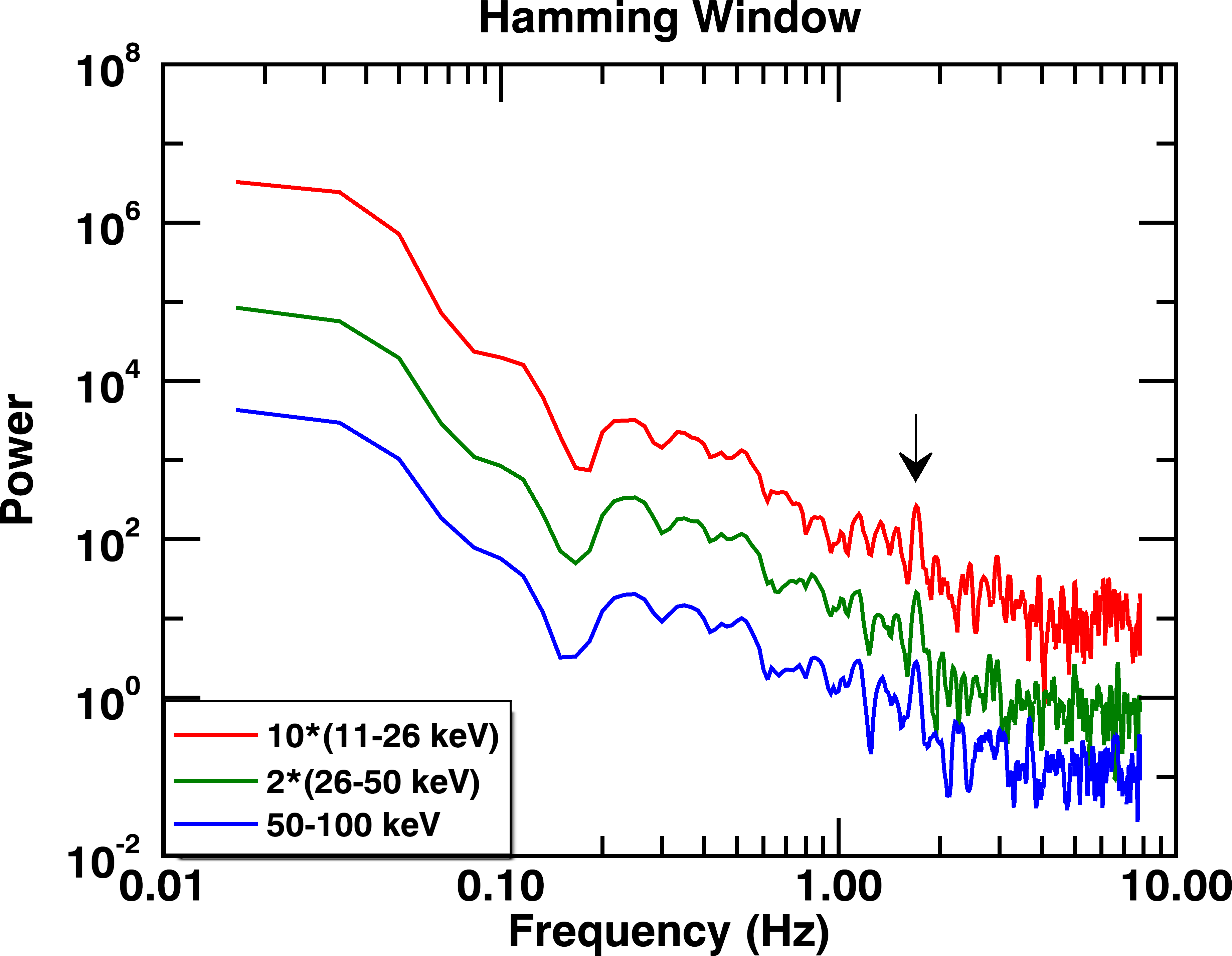}
	\includegraphics[width=0.45\linewidth]{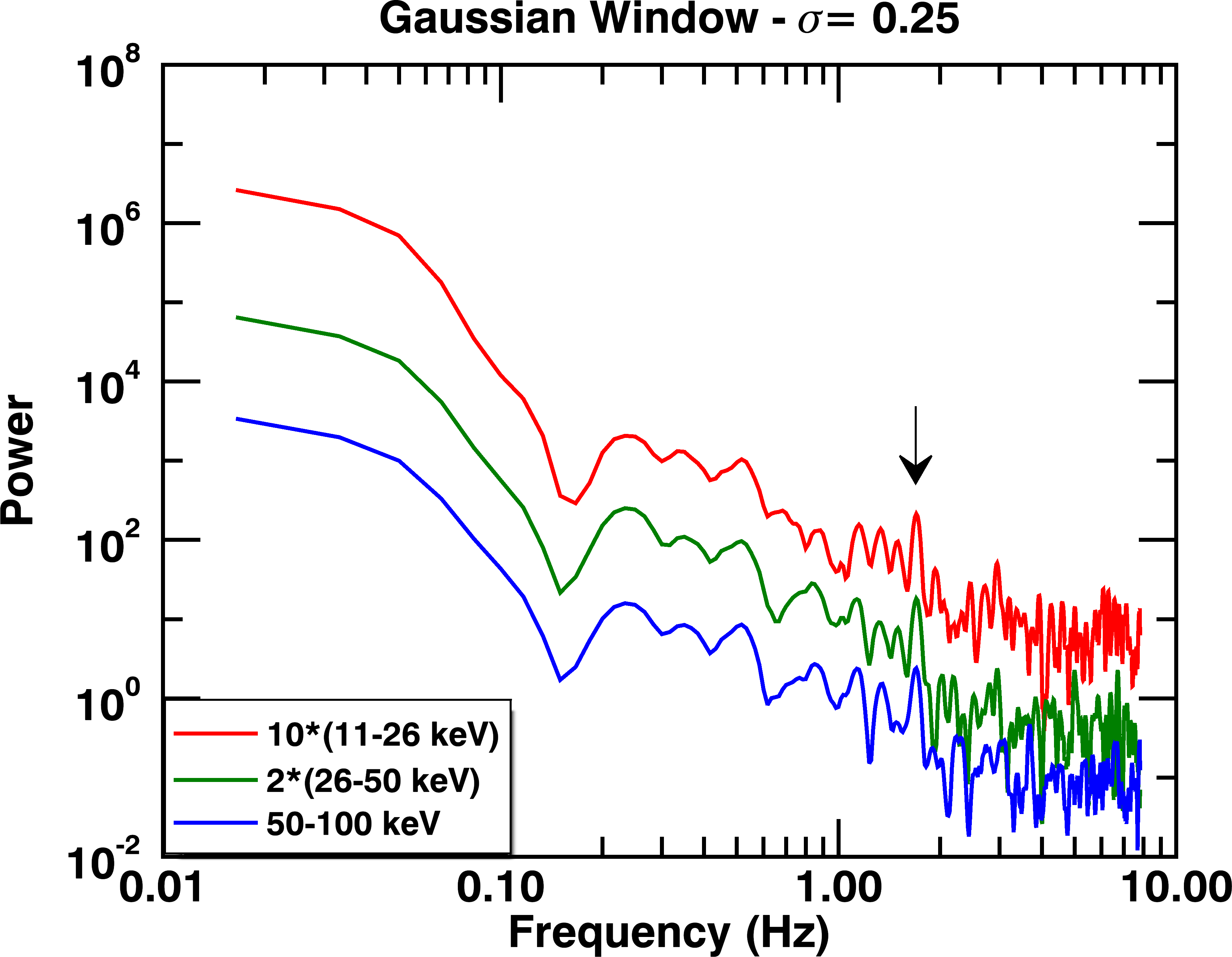}
	\includegraphics[width=0.45\linewidth]{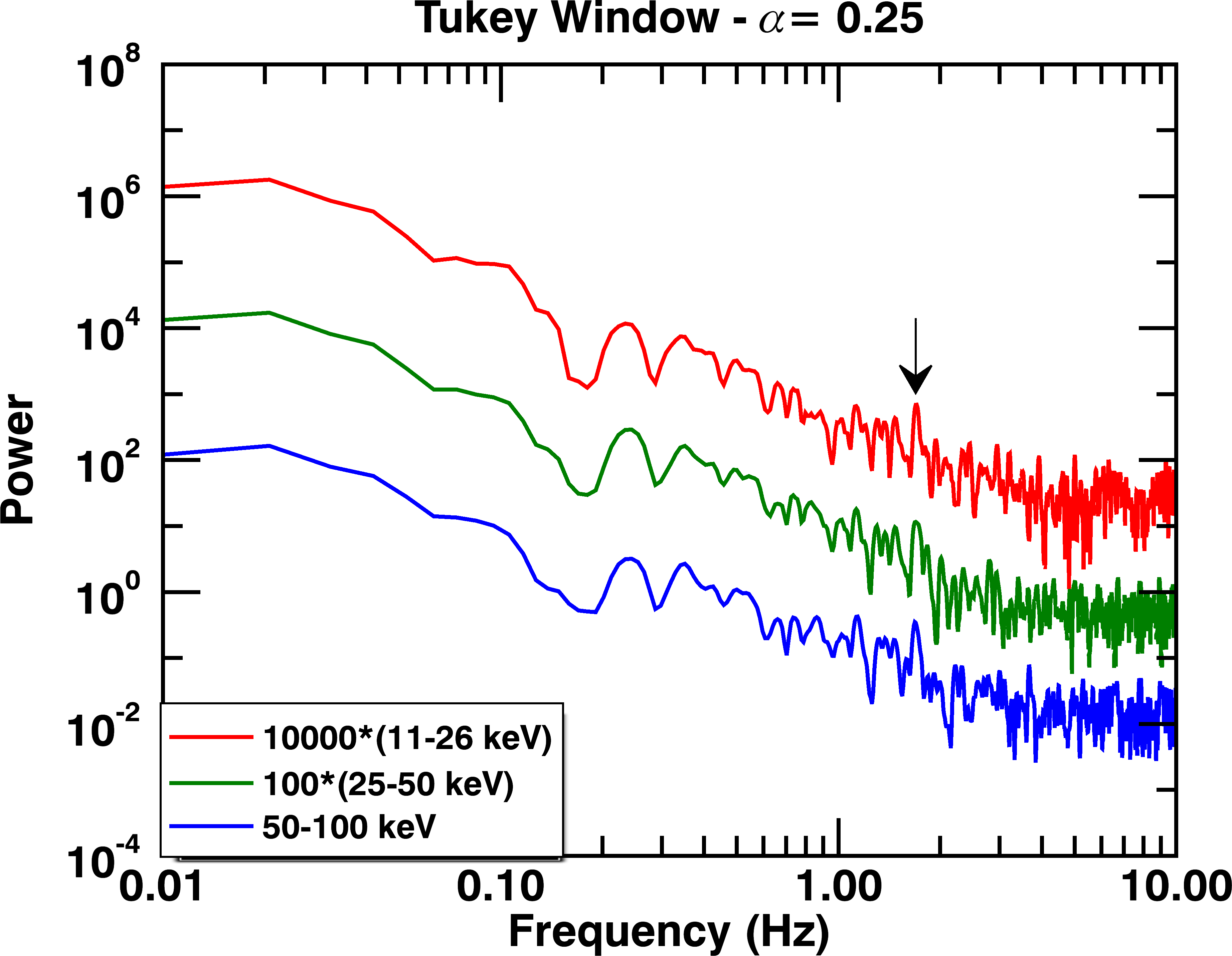}
\caption{ Plots of FFT power spectra for the 3:45:00 to 3:46:00 interval of the SOL2011-08-04 flare utilizing different windowing methods. Curves are multiplied by numerical factors for plotting purposes. For this interval, significant power at the 1.7 $\pm$ 0.1 Hz frequency was found in every power spectrum. A 4-bin smoothing was applied to each of these plots to elucidate strong periodicities. The other intervals in Table \ref{tab:corr} were found to lack any significant periodicity; see Figure \ref{fig:fft_other}. }
\label{fig:fft}
\end{center}
\end{figure*}

\begin{figure}[]
\begin{center}
	\includegraphics[width=0.45\linewidth]{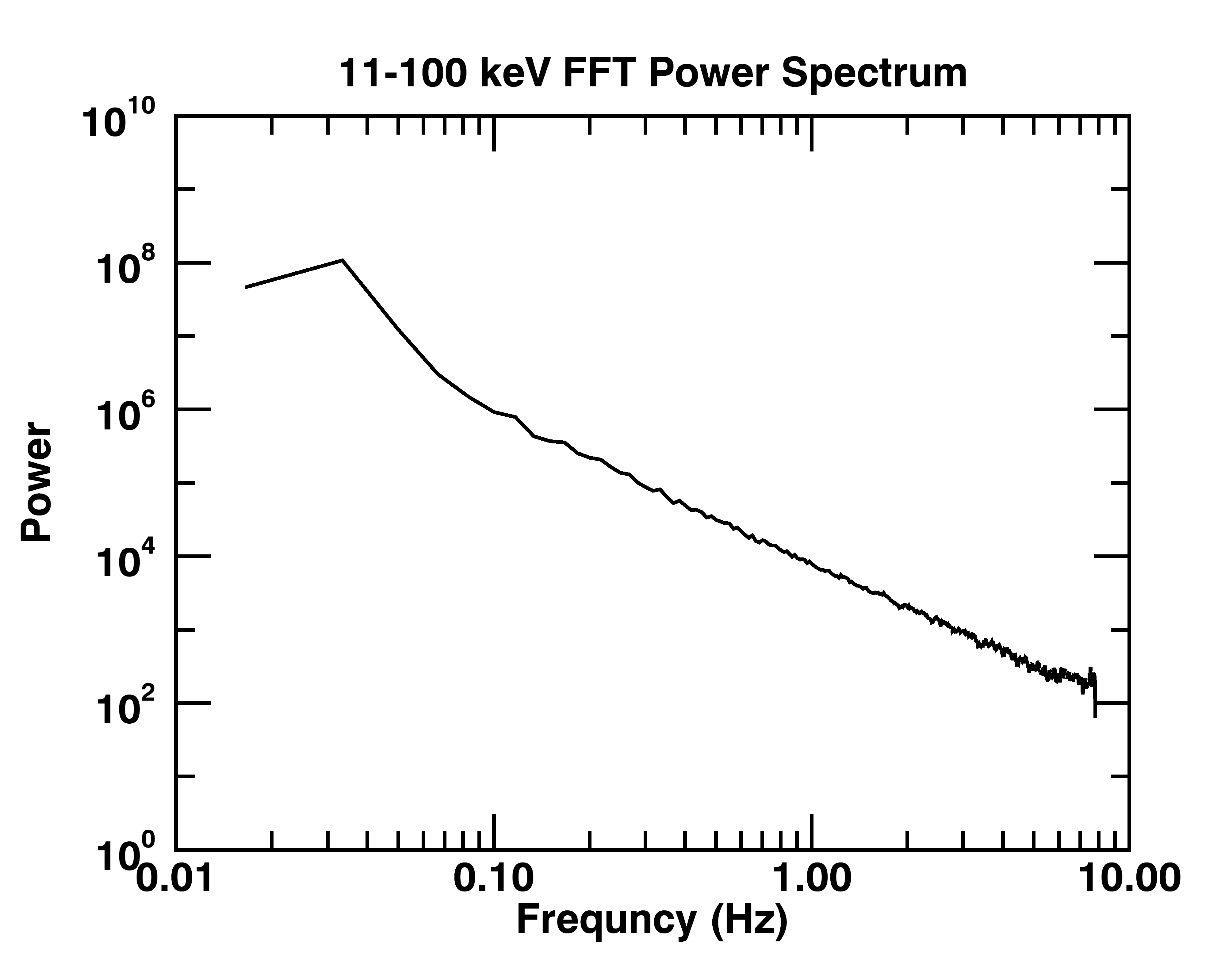}
	\includegraphics[width=0.45\linewidth]{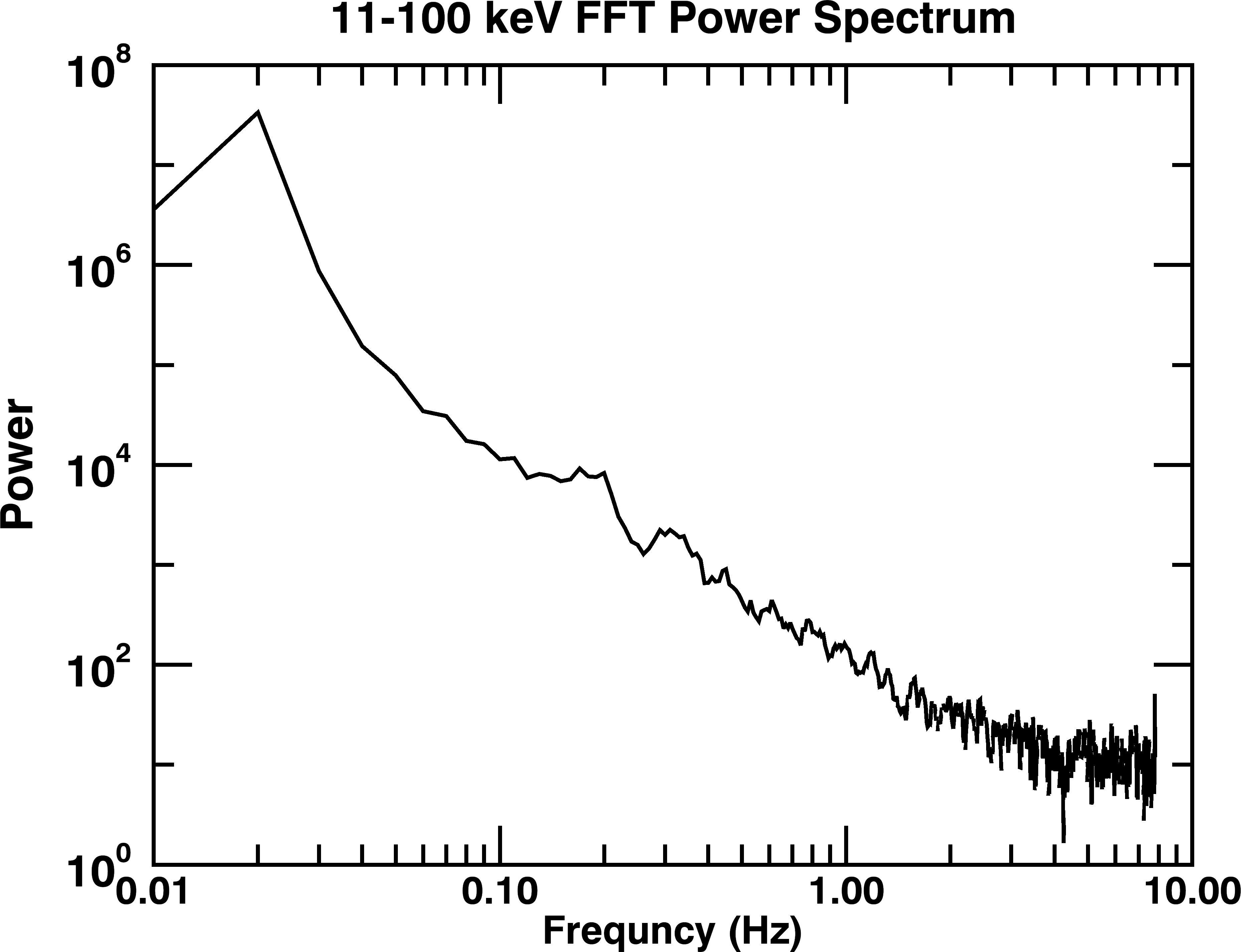}
\caption{ \textbf{Left:} The FFT power spectrum of the 2:07:20 to 2:08:20 interval of the SOL2011-07-30 flare. No significant periodicities were observed. \textbf{Right:} The FFT power spectrum of the 3:49:40 to 3:51:20 interval of the SOL2011-08-04 flare. As in the top image, no significant peridocities were observed.}
\label{fig:fft_other}
\end{center}
\end{figure}

In the 2 chosen flares, subsecond spikes tended to occur in large groupings that last over a span of 60-100 seconds. To assess periodicity within the spikes, a variety of Fourier analysis methods were employed. The techniques applied are a Fast Fourier Transform (FFT), an FFT with a triangular window, an FFT with a Hamming window, an FFT with a Gaussian window, and an FFT with a Tukey window. Applying these windows to FFTs reduces noise due to the abrupt ending of the interval by tapering the interval. A triangle window applies a linear taper, a Gaussian window applies a Gaussian taper, a Hamming window applies a specific case of the Hann filter, a Gauss-like filtering, and a Tukey filter applies a tapered cosine filter. The Gaussian and Tukey windows utilize adjustable parameters, often denoted $\sigma$ and $\alpha$ respectively. These parameters determine the width of the window and 0.25 was used for both parameters in these analyses. These analyses were performed on each energy band independently as well as the summed counts from 11-100 keV. By applying a variety of methods, any peak appearing in all power spectra will have a much higher level of confidence than an individual method would allow. The FFT power spectrum for the SOL2011-08-04 early interval showed a significant peak occuring at $1.7 \pm 0.1$ Hz in all energy bands as seen in Figure \ref{fig:fft}. The FFT power spectra of the other 2 spiking intervals are shown in Figure \ref{fig:fft_other}. No significant periodicites were found in those intervals.

\subsection{Flare Spectroscopy and Pulse Pileup}\label{sec:pileup}
\begin{figure}[]
\begin{center}
	\includegraphics[width=0.45\linewidth]{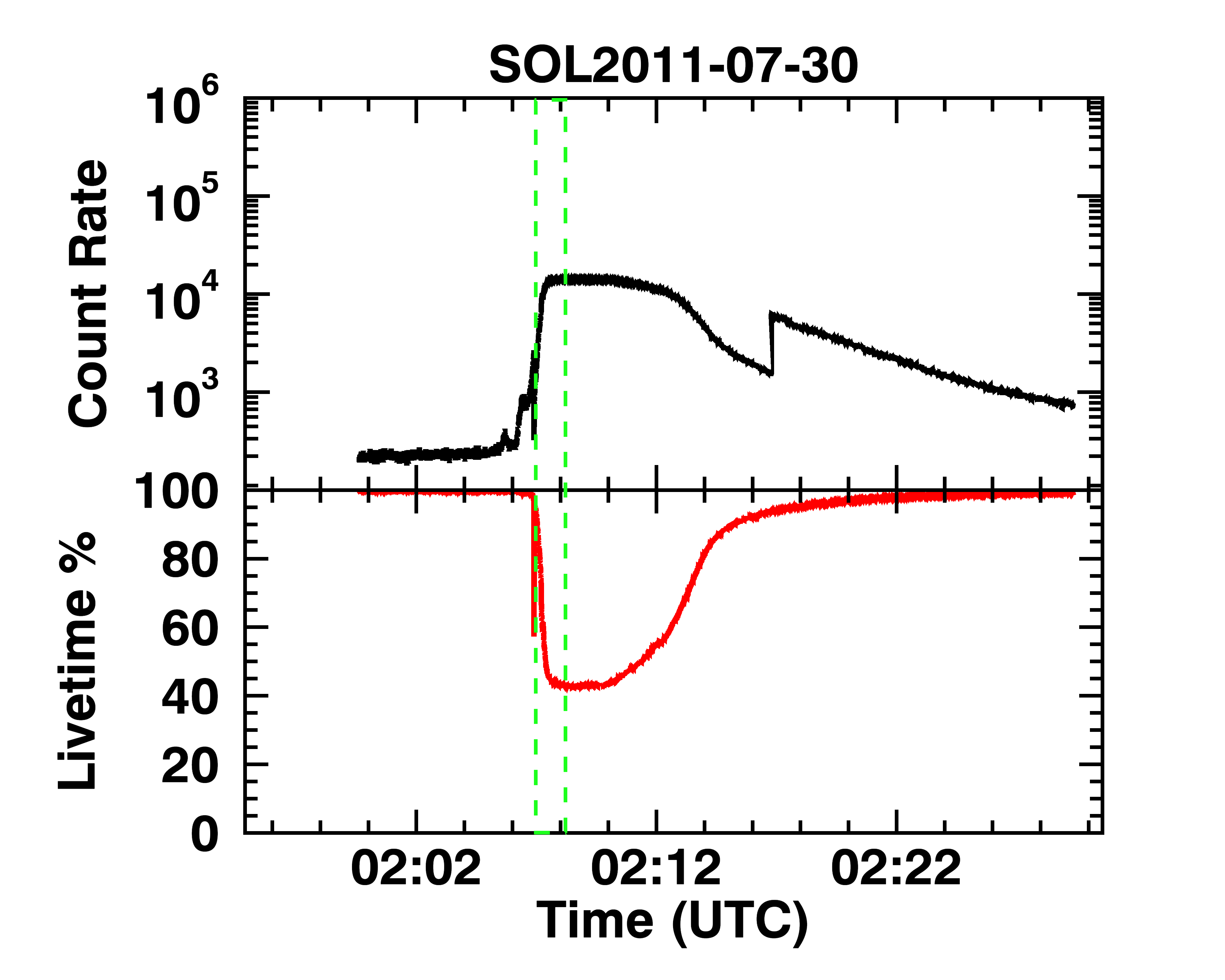}
	\includegraphics[width=0.45\linewidth]{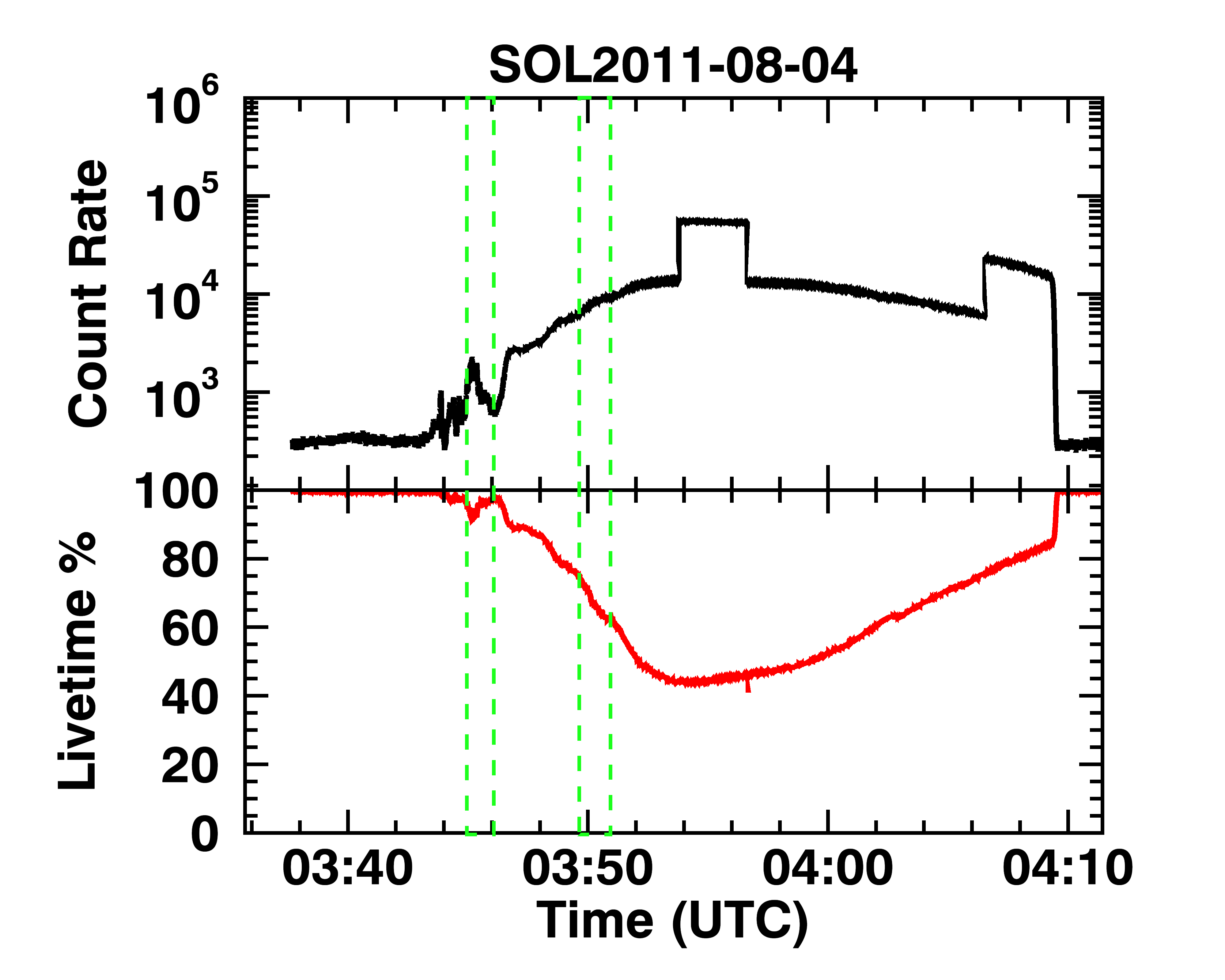}
\caption{ The total count rate and live time percentage of the \textit{Fermi} GBM instrument during the 2 solar flares. The green dashed boxes indicate the intervals in which spikes were found.}
\label{fig:livetime}
\end{center}
\end{figure}

The \textit{Fermi} GBM instrument is designed for detecting nonsolar astrophysical sources. This creates issues when observing solar flares as the flux rates can often exceed the capabilities of the system. When this happens an issue known as pulse pileup occurs. The arrival of 1 or more photons before the shaping and sampling of the previous photon is complete causes photons to ``pileup" into a single measurement of a higher energy. A 0th-order method of diagnosing pulse pileup within a system is to examine the livetime, the period in which the system is ready to receive a signal. As the livetime percentage decreases, the probability of pulse pileup increases. The livetime percentage of GBM during the 2 flares can be seen in Figure \ref{fig:livetime}, reaching a minimum of around 40\% for each flare. While the count rate is below the theoretical maximum of 375,000 events/s \citep{meegan}, pileup is likely occurring. This is further supported by spectral fits of the flares with GBM data as the lower energy component of the spectrum is not well fitted (implying low-energy counts are being piled up into higher energy counts). While pulse pileup will not create time-dependent structures, it can affect our results in a variety of ways. It can make spikes at lower energies appear at higher energies. It can also decrease the amount of lag between energy bands found in Table \ref{tab:corr}, but it cannot increase them. It will also affect any spectral analysis and for that reason we use \textit{RHESSI} data for spectral fitting when possible.

\begin{figure}[]
\begin{center}
	\includegraphics[width=0.45\linewidth]{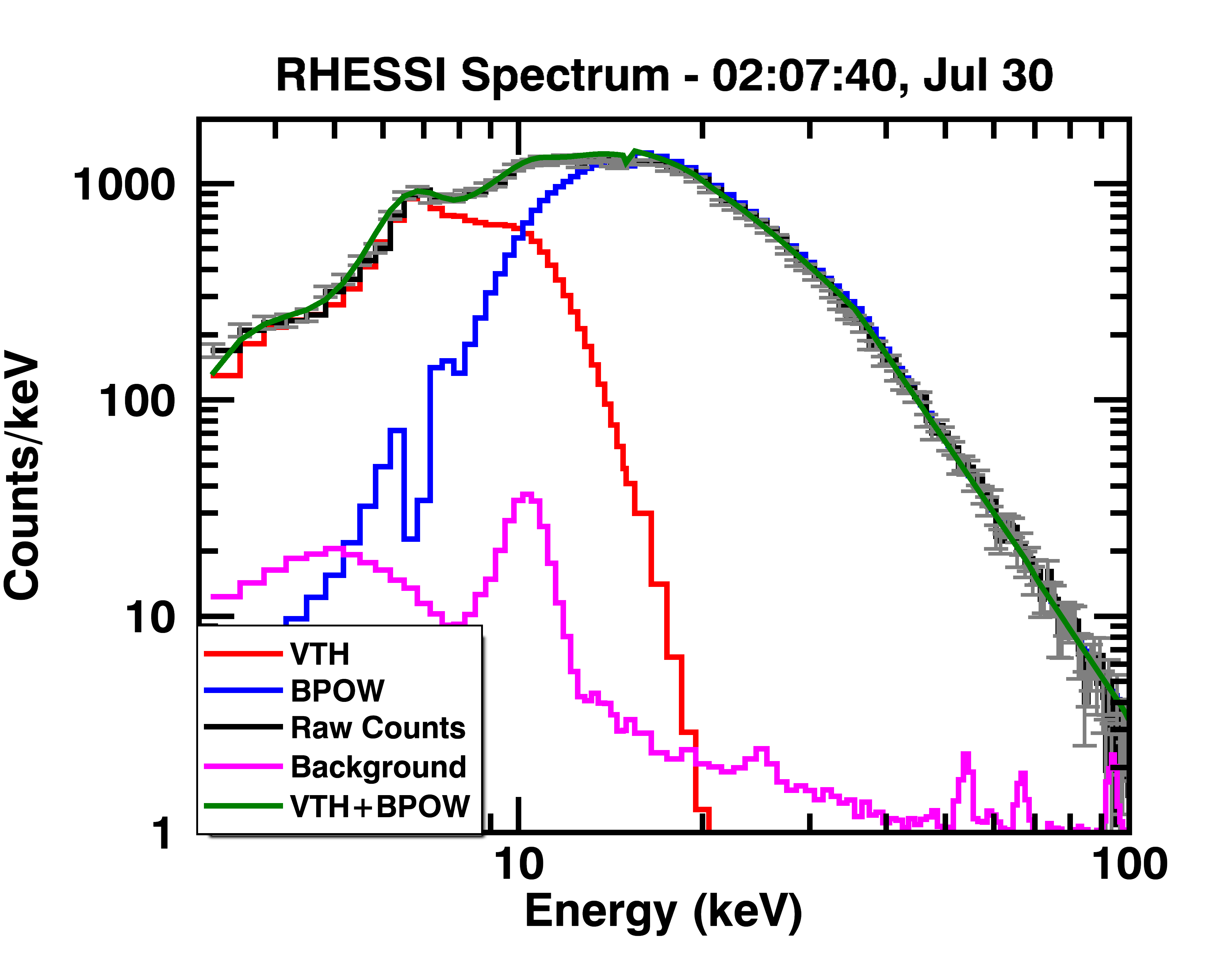}
	\includegraphics[width=0.45\linewidth]{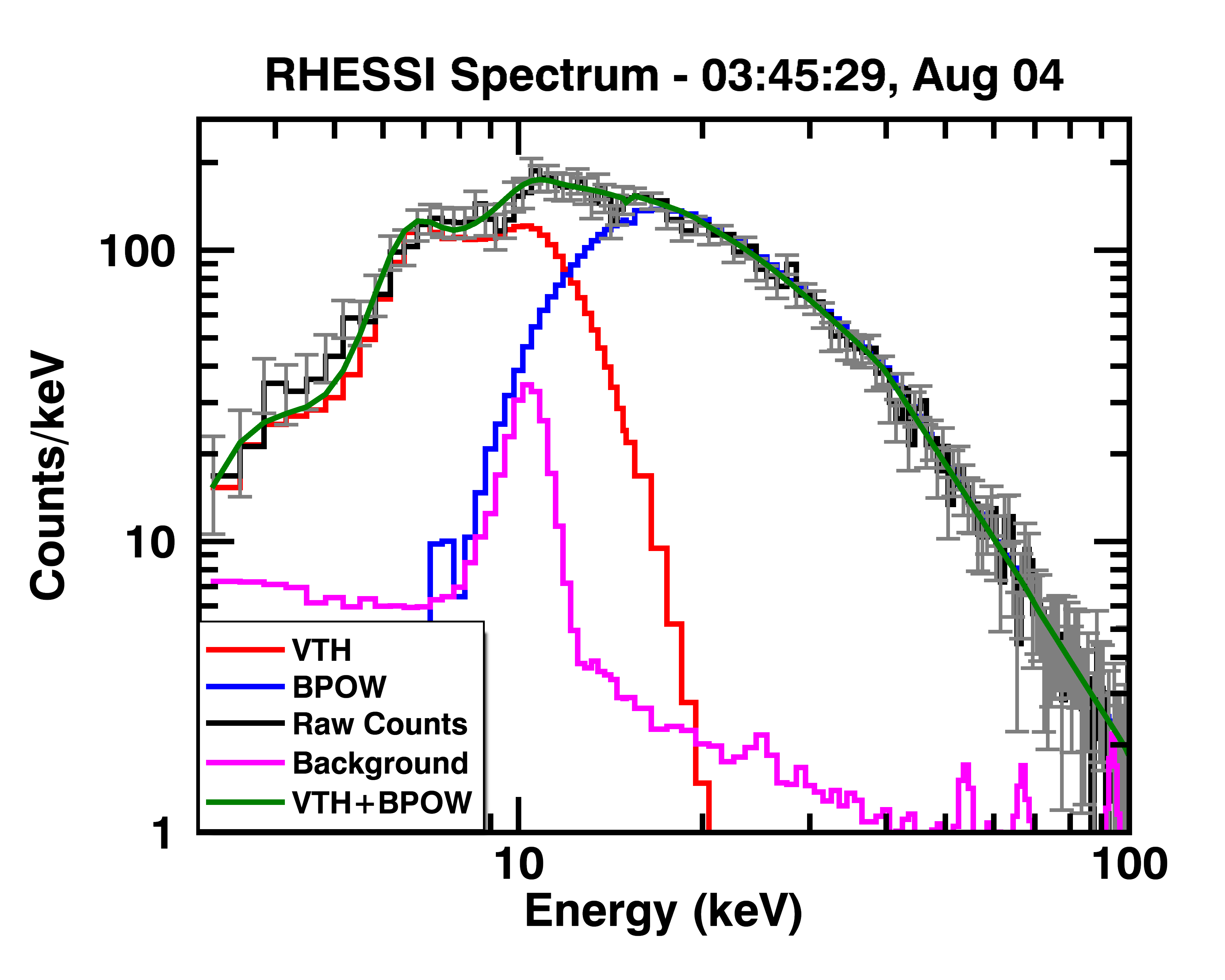}
\caption{ \textit{RHESSI} spectra for 2, 2-second intervals in the spiking periods of both flares. The fit parameters for the left spectral fit were as follows: an emission measure of $1.9\times10^{49} cm^{-3}$, temperature of 10.5 MK, a break energy of 218 keV, and spectral indexes of -3.6 and -9.4 below and above the break, respectively. The fit parameters for the right spectral fit were as follows: an emission measure of $2.5\times10^{47} cm^{-3}$, a tempertaure of 17.6 MK, a break energy of 38.8 keV, and indexes of -2.3 and -3.5 below and above the break, respectively. Both flares demonstrated a turnover in the thermal to nonthermal spectra around 10 keV owing to their occurrence early in the impulsive phase.}
\label{fig:spectra}
\end{center}
\end{figure}

\textit{RHESSI} data are available for the entirety of the SOL2011-07-30 flare but only impulsive phase data are available for the SOL2011-08-04 flare due to spacecraft nighttime. All detected spikes in the SOL2011-08-04 flare were found in the impulsive/peak phase of the flare. Spectral analysis was performed to determine the thermal-nonthermal transition in the spectrum, flare temperatures, and spectral index of the nonthermal flux. Characteristic spectra from these spiking intervals are shown in Figure \ref{fig:spectra}. The SOL2011-07-30 fit parameters showed an emission measure of $1.9\times10^{49} cm^{-3}$, a temperature of 10.5 MK, and fit the nonthermal component with a -3.6 spectral index (the break energy was at 218 keV, effectively giving a single power law fit). The SOL2011-08-04 fit parameters showed an emission measure of $2.5\times10^{47} cm^{-3}$, a temperature of 17.6 MK, a break enegy of 38.8 keV, with spectral indexes of -2.3 and -3.5 below and above the break, respectively. Both showed the thermal to nonthermal transition occurring around 10 keV, largely owing to their occurrence early in the impulsive phase. The SOL2011-07-30 flare had a fitted peak temperature of around 25 MK while the SOL2011-08-04 flare had a fitted peak temperature of around 33 MK.

\begin{figure}[]
\begin{center}
	\includegraphics[scale=0.5]{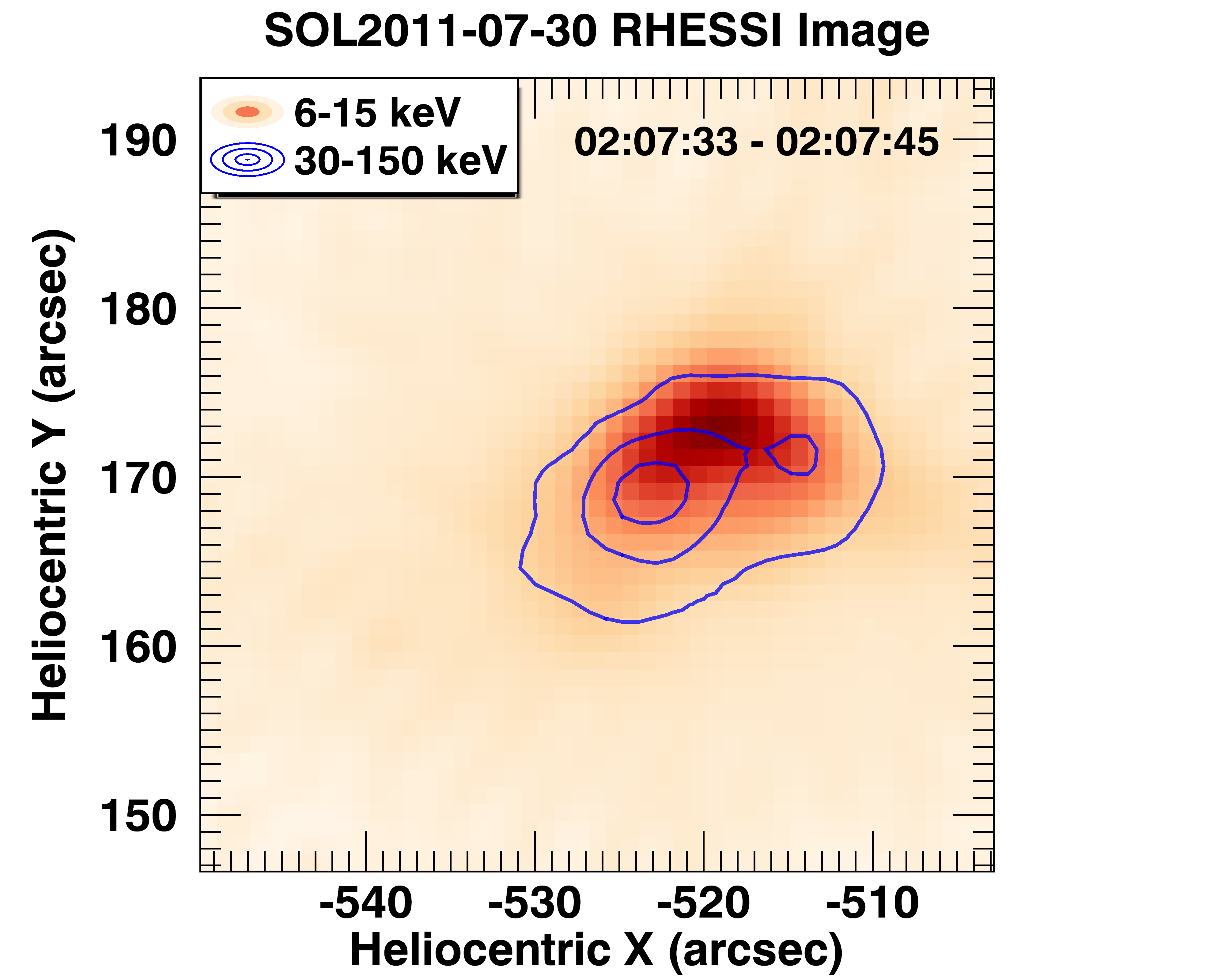}
	\includegraphics[scale=0.5]{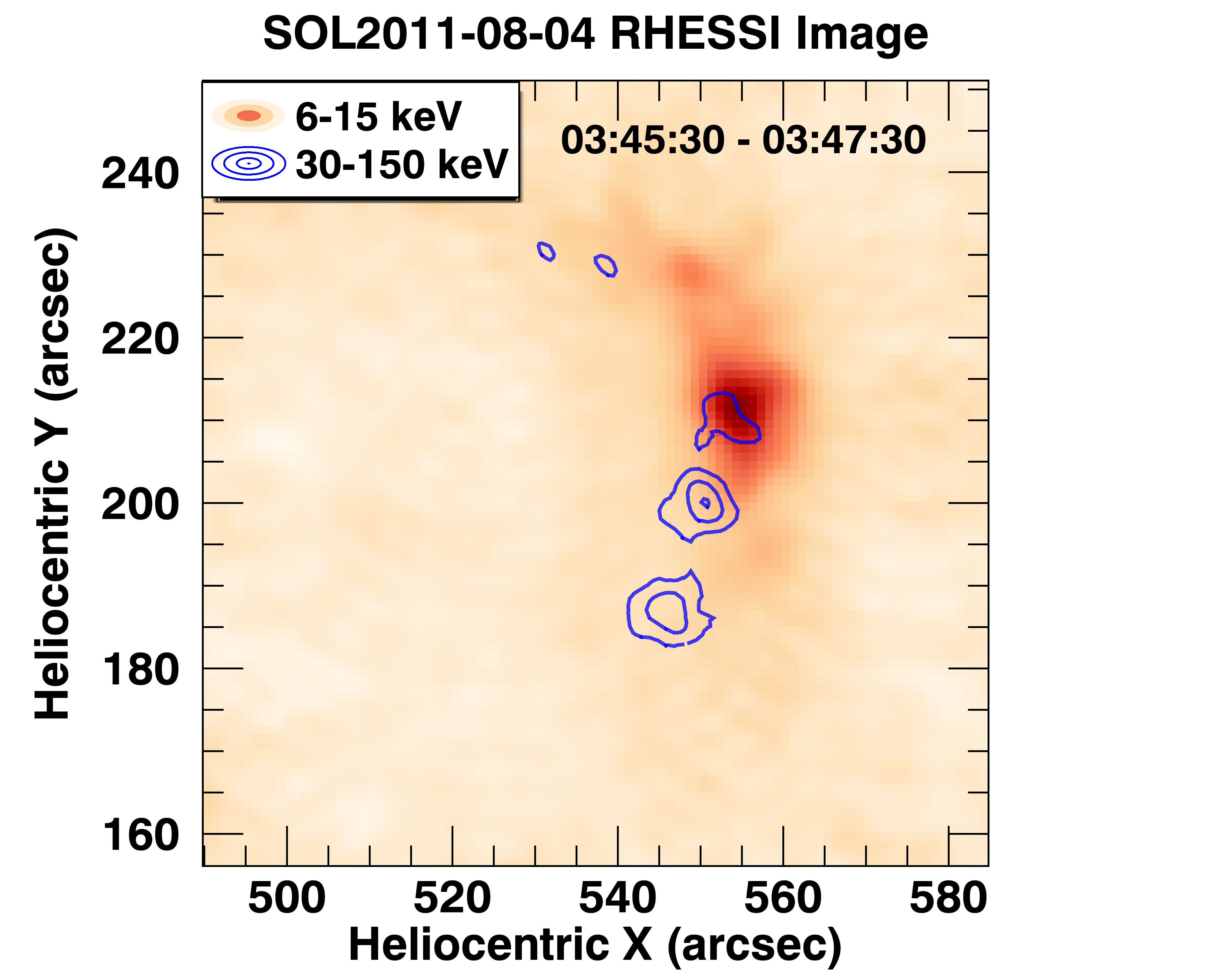}
\caption{ Reconstructed CLEAN \textit{RHESSI} X-ray images of the 2 flares. These images were produced with data from detectors 1-8. Thermal emission in the 6-15 keV range shows a bright, singular source, likely looptop emission. The blue curves show emission in the 30-150 keV range, predominantly coming from 2 spots in each image. These are likely footpoint emission. The contours correspond to emission 50\%, 70\%, and 90\% of the maximum. Note, we do not have spatial information on the spike origins and these images were examined for context.}
\label{fig:images}
\end{center}
\end{figure}

Figure \ref{fig:images} shows CLEAN processed \textit{RHESSI} X-ray images for the 2 flares. These images are from periods in which X-ray spikes were present. Detectors 1-8 were used for both of these images. In both, a clear, singular thermal source is present, likely from loop emission. Both images also show footpoint emission in the 30-150 keV band, though it is less defined in the SOL2011-07-30 flare. The contours correspond to 50\%, 70\%, and 90\% of peak emission. We have no spatial information as for the origin of the spikes, but it is likely they are coming from the regions of higher X-ray emission, as found by \citep{qiu,cheng}.

\section{Discussion}\label{sec:discussion}

The goal of studying these subsecond spikes is to establish a constraint for particle acceleration mechanisms within solar flares. In order to do this, we must have predictables derived from theory to compare against. By calculating the relevant characteristic timescales for these various mechanisms, we can test the validity of acceleration mechanisms such as those described in Section \ref{sec:intro}.

In the case of shock acceleration, it has been shown that should a pair of oblique shock fronts occur within a flare, they can accelerate particles on timescales of $t=L/2u\tan\theta$, with $L$ the distance between the shocks, $u$ the downstream flow velocity with respect to the upstream velocity, and $\theta$ the angle of the shocks \citep{tsuneta}. Tsuneta \& Naito found 0.3-0.6 seconds to be a characteristic timescale for this process. Their study used ``characteristic" flare properties that are similar to the observed properties of these flares. The spacing of the shock fronts however is a free paramater that is not discernible from X-ray observations. The timescales found are congruent with a large portion of the spikes observed, as shown in Figure \ref{fig:dist}. While 2 shock fronts that are particularly close could create sufficiently fast acceleration, ultimately this fails to account for the numerous faster spikes. Shock acceleration also suffers from the issue of requiring a pre-energized population of electrons in order to accelerate efficiently and it is unclear if there exist enough electrons in the higher end of the thermal distribution to supply this population. A faster mechanism, responsible for the shorter time bursts we observe that then feeds into a shock acceleration mechanism, is plausible. With these factors in mind we can conclude that direct, oblique shock acceleration, while potentially present and responsible for a large number of the spikes, is not the only mechanism in these flares. Acceleration can also occur in collapsing magnetic traps \citep[e.g.][]{somov,drake}. In these scenarios the acceleration timescales are directly proprotional to the rate of collapse and size of these traps. [Simulations of magnetic islands have often shown them to follow a roughly power-law distribution \citep{guidoni}, which could explain some of the characteristics observed, such as the -1.2 $\pm$ 0.3 power law index found in the SOL2011-08-04 spike duration distribution, shown in Figure \ref{fig:dist}]. A more in-depth examination of the existence and parameters of magnetic islands within these flares is needed for comparison.

Second-order Fermi acceleration (also known as stochastic acceleration) is the acceleration of particles via turbulent motions within the solar flare plasma. Cascading MHD turbulence involves moving magnetic compressions resulting from MHD waves. As the cascading occurs, particles undergo Cherenkov resonance. This resonance diffuses particles across momentum space, setting up additional resonances with the cascading MHD waves \citep{chen, zharkova}. Determining the characteristic timescales for this mechanism is difficult as it is highly dependent upon physical parameters such as the momentum and pitch angle diffusion coefficients, the turbulence diffusion coefficient, and total energy distribution of the accelerated particles. These can all vary between particular flares. \cite{chen} calculated acceleration timescales of 0.96s for an X3.9 flare and 0.45s for an M2.3 flare by applying a ``Leaky Box" model where particles diffused out of a magnetic trap via loss-cone physics. The diffusion coefficients and particle energy distribution were calculated via the Fokker-Planck equation and the resultant X-ray spectra. \cite{alty} also studied the SOL2011-08-04 flare using observations in multiple wavelengths, primarily focusiong on Nobeyama Radioheliograph data. They found acceleration timescales on the order of 50 ms or shorter and concluded it was unlikely that turbulent stochastic models were sufficient. They instead offered super-Dreicer electric field acceleration \citep{priest2000} as the most plausible option. Dreicer field acceleration in solar flares is still not well formulated but may result from current sheet development near the reconnection region \citep{alty}.

Differences in the timing of these spikes across energies can reveal time of flight effects such as particle trapping or it can indicate a ``slow" acceleration mechanism that struggles to accelerate to higher energies. Table \ref{tab:corr} showed a systematic peaking in the higher energy bands first. While the uncertainties are relatively large, all results are inconsistent with 0. Due to difficulties arising from projection effects, we assumed a ``typical" half-loop length of 10 Mm \citep{priest} and direct flight path, a 20 and an 80 keV electron would have an arrival time difference of approximately 0.06 seconds, on order of the binning of \textit{Fermi} GBM CTIME data. The short lag times observed indicate an acceleration mechanism that can accelerate particles to higher energies rather quickly. This is further supported by the findings of \cite{alty} which indicated acceleration timescales of 50 ms or shorter for one of our flares.

The 1.7 $\pm$ 0.1 Hz quasi-periodicity found in the first spiking interval of the SOL2011-08-04 flare is interesting. If the periodic behavior is due to particle motion, then there is relation between a typical distance traveled, the period, and the particle velocity. We calculated that the corresponding loop lengths for 20, 40, 60 and 80 keV electrons at this frequency to be 48, 65, 76, and 88 Mm respectively. This was done by converting the electron energies into velocities using a relativistic equation and then assuming mirroring at each footpoint corresponded to the bursts of emission. There is no reason to believe that particles of different energies would see loop lengths of different sizes. Thus, becuase we observe the frequency in all energy bands, it is unlikely to be due purely to particle motion. Instead, a possibility is that the acceleration mechanism itself features a periodicity. If the energy release mechanism is oscillatory, the energization of particles would would follow a similar pattern. A common example of such a mechanism is the ``dripping" model which involves a constant flow of energy into the reconnection region. The energy builds up until it overcomes an effective ``tension" force \citep{nakariakov}. However the period of these processes in flares tends to be on the order of minutes and thus explains the quasi-periodic pulsations (QPPs) observed on longer timescales \citep{asai}. A more plausible explanation is the existence of an oscillating plasma structure near the reconnection site. Such a structure has been shown as capable of creating periodic, fast magnetoacoustic waves which would generate ``very localised and sharp spikes of electric current density" \citep{nakariakov}.

\section{Conclusion}

Particle acceleration in solar flares remains an ongoing area of research as questions on flare energetics, scaling with flare size, and reconnection mechanisms remain unanswered. Fast-time domain analysis of solar flare X-ray flux has been an oft-ignored method of diagnosing these acceleration mechanisms due to the lack of direct solar flare X-ray time profiles. The \textit{Fermi} GBM is an instrument with adequate energy range and time resolution to probe subsecond X-ray phenomenon. We examined two M9.3 class solar flares occurring on July 30 and August 04, 2011 as a case study for measuring subsecond spikes in solar flare X-ray flux. By applying a smoothe subtraction to the light curves, subsecond spikes were identified in the residuals via a stringent selection criteria. A Gaussian Mixture Model was fit to the spikes using an expectation maximization algorithm in conjunction with a Bayesian Information Criterion in order to conservatively estimate the number of spikes. We then transferred the number of Gaussians and their means to a least squares fitting algorithm and fit the spikes in raw count space in conjunction with a quadratic background function, ultimately providing the spike durations and their peak time per energy band. We also applied methods to determine the quasi-periodicity and spike lag across energies.

This study is a case study of subsecond spikes in \textit{Fermi} Gamma-ray Burst Monitor solar flare data as well as the establishment of a comprehensive method of identifying and analyzing those spikes, built upon the work of previous studies \citep[e.g.][]{kiplinger1983,kiplinger1984, qiu, cheng}. The short time domain of solar flare X-ray flux has been an underexplored area for the last 20 years, largely owing to \textit{RHESSI}'s modulation of time signals on the order of 4 seconds. The next generation of high energy solar X-ray observatories has yet been finalized, but should include subsecond resolution as an aspect of their design. One such instrument, the IMpulsive Phase Rapid Energetic Solar Spectrometer (\textit{IMPRESS}), is in development via a collaboration between the University of Minnesota, Montana State University, University of California-Santa Cruz, and Southwest Resarch Institute. This 3U CubeSat will boast a full-sun spectrometer with an energy range of 4-100 keV and provide hard X-ray (8-100 keV) histograms on $\sim$32 Hz cadence. Its capabilities allow for both spectroscopy while still providing high resolution time profiles without the significant pileup issues \textit{Fermi} suffers.

\section{Acknowledgements}
We would like to thank Laura Hayes and Andy Inglis for their input on the time profile analysis of this work and Richard Schwartz for his input on the \textit{Fermi} GBM data products. We would also like to thank NASA for supporting this research through the NASA Earth and Space Science Fellowship program (grant 80NSSC18K1118) and the NASA Drive Science Center (grant 80NSSC20K0627) and the NSF for supporting this research through the NSF Faculty Development Grant (grant AGS-1429512).

\end{document}